\documentclass[a4paper,12pt]{article}
\usepackage[utf8]{inputenc}
\usepackage{multicol}
\usepackage[utf8]{inputenc}
\usepackage{amsmath}
\usepackage{amssymb}

\usepackage{subcaption} 
\usepackage{float} 
\usepackage{graphicx}
\usepackage[dvipsnames]{xcolor}
\usepackage{graphicx,times}            
\usepackage{amssymb,amsmath}

\DeclareUnicodeCharacter{2212}{\textendash}
\usepackage{graphicx}% Include figure files
        \usepackage{dcolumn}% Align table columns on decimal point
        \usepackage{bm}% bold math

        \usepackage{lipsum}
        \usepackage{url}
        \usepackage{hyperref}
        \usepackage{multirow}
        \usepackage[dvipsnames]{xcolor}
        \usepackage{comment}

\title{\normalsize\bf Charged black holes in Eddington-inspired Born-Infeld gravity: An in-depth analysis\\ of the structure of spacetime geometry}

        \author{\normalsize\sc Muhammed Shafeeque\footnote{m.shafeeque@iitg.ac.in}\ \  and Malay K. Nandy\footnote{mknandy@iitg.ac.in}\\
        \small\em Department of Physics, Indian Institute of Technology Guwahati, Guwahati 781 039, India.}

        \date{\small (December 10, 2024)}

\begin{document}
\maketitle

\begin{abstract}
In this paper, we focus upon the behaviour of spacetime of charged black holes described by Eddington-inspired Born-Infeld (EiBI) gravity. With a static and spherically symmetric metric, we solve the ensuing field equations obtained from the EiBI-Maxwell action in the Palatini formalism. Consequently we carry out, for the first time, an in-depth analysis of the structure of spacetime geometry in several regions of the charged EiBI black hole. In particular, we consider the analytical behaviours of the metric coefficients and the Kretschmann scalar by probing their asymptotic nature {\em analytically} in different regions of the black hole spacetime, such as, near the center,  in the intermediate region, and near the horizon, for both positive and negative EiBI coupling. These analyses give a thorough understanding of the nature of spacetime of EiBI-Maxwell black holes. In order to aide our understanding further, we solve the EiBI-Maxwell field equation numerically with different values of the parameters involved. We find close agreement between the analytical behaviours and those obtained from numerical integration of the EiBI-Maxwell field equation.

\end{abstract}

\tableofcontents

%\section*{abstract}

\section{Introduction}\label{sec_intro}
Einstein's pioneering formulation of general relativity (GR) is the most successful theory of gravity, having passed various tests in weak gravitational scenarios of the solar system \cite{eddington1923mathematical,Shapiro1964, Adelberger_etal_2003_rev, Will2014_rev}. With the advancement of technology, various other predictions of GR, such as production of gravitational waves in binary collapse \cite{Ligo_GW170817}, and the existence of black hole at the centre of the galaxy\cite{Akiyama_2019_M87bh,SagittariusA_i,SagittariusA_ii,SagittariusA_iii}, have also been confirmed. However, GR predicts the existence of an essential singularity in the black hole and in the big-bang cosmology of the universe  \cite{Penrose_1965,Penrose:1969pc,Hawking1972,mtw1973,Hawking1976,Penrose:1980ge,wheeler1981,susskind2008black,Starobinsky_1981vz,inflation_1}. This shows that GR is incapable of explaining the physics in strong gravitational situations. This dilemma warrants the search for a new theory of gravity which has been a major goal in research in modern times. There have been various modifications proposed \cite{Brans_Dicke_1961,Bergmann1968,Nordtvedt1970,Damour1992,STAROBINSKY1980,CAPOZZIELLO,Deser_1998,Harko2011} in order to explain the strong gravity regime.

Ever since the formulation of Einstein's general relativity, physicists have continually sought alternative theories of gravity. One such theory, proposed by Eddington at the dawn of the last century \cite{eddington1923mathematical}, has recently gained renewed attention.  Eddington's theory is incomplete since matter was not coupled to the gravity in that framework. A modification of Eddington's theory, accommodating matter field has been coined as Eddington-inspired Born-Infeld (EiBI) gravity \cite{EiBI_2010}.  This theory integrates  Eddington's original idea with that of Born-Infeld electrodynamics \cite{Born_infeld_1934} into a unified theory.

The EiBI gravity is a Palatini-type theory \cite{Palatini1919}, distinguishing itself from general relativity by treating the affine connection $\Gamma^\mu_{\nu\rho}$ and metric tensor  $g_{\mu\nu}$ as independent geometric entities. Consequently, the EiBI action is varied with respect to both $\Gamma^\mu_{\nu\rho}$ and $g_{\mu\nu}$ independently, resulting in two equations of motion that couple matter with gravity.

EiBI gravity differs from general relativity in the presence of matter. In the absence of matter, EiBI reduces to general relativity. The strength of gravity in the EiBI theory is determined by a coupling parameter $\kappa$. For small values of $\kappa R_{\mu\nu}$, EiBI gravity reduces general relativity, while for large values of $\kappa R_{\mu\nu}$, it approaches Eddington gravity. When the matter field vanishes, the EiBI action reduces to the Eddington action, which is equivalent to the Einstein-Hilbert action, yielding general relativity \cite{EiBI_2010}.

The EiBI theory of gravity offers a promising avenue for further exploration.  The distinction from general relativity becomes particularly relevant in the strong gravity regime, especially in the presence of matter, as discusse above. Under such conditions, the Poisson equation is modified in EiBI for strong gravitational fields or dense matter\cite{pani_etal_2011}. This characteristic is of paramount importance for studying compact objects like black holes \cite{EiBI_2010,Sotani_Miyamoto_2014,Sotani_2015,Wei2015,Avelino_2016,Jana_etal_2018,chen_chen_2018} and neutron stars \cite{pani_etal_2011,Pani_Sotiriou_2012,Harko_etal_2013,kim_2014},  exhibiting deviations from general relativity when interacting with dense matter.

The EiBI gravity further offers a promising avenue for addressing the cosmological singularity problem. Research has shown that a universe filled with regular matter in EiBI has a maximum finite density (or minimum length) limit, preventing the formation of the cosmological singularity \cite{Avelino_Ferreira_2012}. Additionally, depending on the specific parameter values, EiBI can exhibit a regular bounce from contraction to expansion, oscillating between two finite density values before transitioning to a GR-like expansion phase \cite{Scargill_etal_2012}. Within the cosmological framework, EiBI has been extensively studied in relation to inflation \cite{Cho_etal_2013}, linear perturbations \cite{Yang_etal_2013}, scalar perturbations \cite{Cho2015,Li_Wei_2017}, and large-scale structure formation \cite{Du_etal_2014}.

Resolution of the cosmological singularity has prompted investigations into the collapse of matter. Studies of non-relativistic \cite{pani_etal_2011} and relativistic \cite{pani_etal_2011,pani_etal_2012} dust collapse in EiBI have revealed that singularities do not form in EiBI. While the positive coupling parameter $\kappa$ occurring in the EiBI theory prevents singularities in general, in the case of relativistic collapse, a constraint on the parameter ($\kappa$) emerges, making the formation of black holes still possible \cite{pani_etal_2012}. Additionally, analysis of $2+1$ dimensional collapse in EiBI \cite{Shaikh_joshi_2018} has shown that singularities, when they do form, are typically naked except within a specific range of parameter values.

EiBI deviates from general relativity primarily in the presence of matter, and the simplest matter configuration within a black hole is an electric charge \cite{Sotani_Miyamoto_2014,Wei2015,Sotani_Miyamoto_2015}. Analysis of charged black hole solutions in EiBI showed an asymptotically flat solution that will approaches the de Sitter form as $r\rightarrow\infty$. For the coupling parameter $\kappa>0$ as well as $\kappa<0$, there exist an event horizon with singularities in a certain parameter range. Moreover, for certain parameter ranges, there is no bound on the black hole charge in contrast with Reissner-Nordstr\"om solution in general relativity. In comparison with general relativity, the horizon radius is smaller (larger) for the coupling parameter  $\kappa>0 $ ($\kappa<0$) \cite{Sotani_Miyamoto_2014}.

The black hole solutions in EiBI \cite{Sotani_Miyamoto_2014} clearly demonstrate  deviations from GR, especially in the vicinity of the black hole. Extending the solutions interior to the event horizon reveals two singularities: at $r=0$ and at $r=\sqrt{\sqrt{\kappa}Q}$, for a black hole with charge $Q$. This behavior differs significantly from that of general relativity.

Although charged black holes in the EiBI framework have been extensively studied, for a deeper understanding of the nature of spacetime of the black hole requires further studies by means of analytical considerations.

In this work, we therefore consider a charged black hole in the EiBI framework. With a static spherically symmetric metric, we first obtain the field equations from the EiBI action minimally coupled to the Maxwell action. This leads to a differential equation for the metric coefficient that contains the charge $Q$ and the EiBI coupling $\kappa$ giving it a very intricate form. With this, we probe the nature of the metric coefficient by {\em analytically} studying its behaviour in different regions of the black hole spacetime. Consequently, we obtain the analytical nature of the metric coefficients and the Kretschmann scalar in the region near the center, in the intermediate region, in the region near the horizon, for both positive and negative EiBI coupling ($\kappa>0$ and $\kappa<0$). This analytical study gives a thorough understanding of the nature of spacetime of the EiBI-Maxwell black hole.

In order to aide our understanding further, we solve the EiBI-Maxwell field equation numerically with different values of the parameters involved. We find close agreement between the analytical behaviours and those obtained from numerical integration of the EiBI-Maxwell field equation.

The rest of the paper is organised as follows. In Section \ref{sec_model}, we consider the EiBI action with a minimally coupled Maxwell field and obtain the field equations in the Palatini formalism. In Section \ref{sec_asymp}, we carry out an in-depth analysis of the nature of spacetime in several regions of the charged EiBI black hole. There, we obtain the analytical behaviours of the metric coefficients and the Kretschmann scalar by probing their asymptotic nature analytically in different regions of the black hole spacetime for both positive and negative EiBI coupling. In Section \ref{sec_num}, we solve the EiBI-Maxwell field equation numerically, where we illustrate the radial profiles of the metric coefficients and the Kretschmann scalar for different values of the parameters involved. Finally, we conclude the paper with a discussion in Section \ref{sec_disc_con}.

\section{EiBI-Maxwell action and field equations}\label{sec_model}

Eddington proposed \cite{eddington1923mathematical} that the relevant field in free (de Sitter) space is the affine connection $\Gamma^\mu_{\nu\rho}$ and that the correct action should be $S_{\rm Edd}=2\kappa\int d^4x\sqrt{|R|}$, where $\kappa$ is a constant having the dimension of $\Lambda^{-1}$, where $\Lambda$ is the cosmological constant.

In the EiBI model, the Eddington action is modified in analogy with the Born-Infeld model of electrodynamics \cite{Born_infeld_1934}. The EiBI action for any general astrophysical system \cite{EiBI_2010} is given by

\begin{equation}\label{eq_action}
 \begin{aligned}S(g,\varGamma,\Psi)= & \frac{1}{\kappa}\int d^{4}x\left(\sqrt{-|g_{\mu\nu}+\kappa R_{\mu\nu}(\Gamma)|}-\lambda\sqrt{-g}\right)\\
 & +\int d^{4}x\sqrt{-g}\mathcal{L}_{m}(g,\Psi),
\end{aligned}
\end{equation}
where $\lambda=1+\kappa\Lambda$, $g_{\mu\nu}$ is the metric tensor, $\mathcal{L}_{m}(g,\Psi)$ is the matter Lagrangian (a function of the metric tensor $g_{\mu\nu}$ and the matter field $\Psi$). It is important to note that the affine connection $\Gamma^\mu_{\nu\rho}$ and Ricci tensor $R_{\mu\nu}(\Gamma)$ are independent of the metric tensor $g_{\mu\nu}$ in Palatini formalism \cite{Palatini1919}. Therefore, we vary the action separately with respect to  $g_{\mu\nu}$ and $\Gamma^\mu_{\nu\rho}$, resulting in two distinct equations of motion,
\begin{equation}\label{eq_eom1}
 q_{\mu\nu}=g_{\mu\nu}+\kappa R_{\mu\nu}(\Gamma)
\end{equation}
and
\begin{equation}\label{eq_eom2}
 \sqrt{-q}q^{\mu\nu}=\lambda\sqrt{-g}g^{\mu\nu}-\kappa\sqrt{-g}T^{\mu\nu},
\end{equation}
where $q_{\mu\nu}$ is an auxiliary metric with determinant $q$  and the affine connection is defined as
\begin{equation}
 \Gamma^{\mu}_{\alpha\beta}=\frac{q^{\mu\nu}}{2}\left(-\partial_\nu q_{\alpha\beta}+\partial_\beta q_{\alpha\nu}+\partial_\alpha q_{\nu\beta}\right).\label{eq_affine}
\end{equation}

For black holes having an electric charge $Q$, the matter Lagrangian is

\begin{equation}
 \mathcal{L}_{m}=-\frac{1}{2}F_{\mu\nu}F^{\mu\nu},
\end{equation}
where $F_{\mu\nu}=\partial_\mu A_\nu-\partial_\nu A_\mu$, and $A_\mu$ is the electromagnetic vector potential.  The corresponding Maxwell's equation and the energy-momentum tensor are given by
\begin{equation}\label{eq_maxwell}
 \nabla_{\mu}F^{\mu\nu}=0
\end{equation}
and
\begin{equation}\label{eq_em_tensor}
 T_{\mu\nu}=2F_{\mu}^{\ \sigma}F_{\nu\sigma}-\frac{g_{\mu\nu}}{2}F_{\alpha\beta}F^{\alpha\beta}.
\end{equation}
It is important to note that the equation of motion for the electromagnetic field (\ref{eq_maxwell}) depends only on the metric $g_{\mu\nu}$ and independent of the auxiliary metric $q_{\mu\nu}$.

For a static and spherically symmetric spacetime around the black hole, we consider the metric $g_{\mu\nu}$ to be given by

\begin{equation}\label{eq_metric}
 ds^{2}=-f(r)h^{2}(r)dt^{2}+\frac{1}{f(r)}dr^{2}+r^{2}\left(d\theta^{2}+\sin^{2}\theta d\varphi^{2}\right),
\end{equation}
 and the auxiliary metric $q_{\mu\nu}$ as
\begin{equation}\label{eq_auxi_metric}
ds_{q}^{2}=-A(r)B^{2}(r)dt^{2}+\frac{1}{A(r)}dr^{2}+C^{2}(r)\left(d\theta^{2}+\sin^{2}\theta d\varphi^{2}\right).
\end{equation}

For the electromagnetic vector potential, we consider the ansatz $A_\mu=\left(V(r), 0, 0, 0\right)$, where $V(r)$ is the electrostatic potential. Using Maxwell's equation (\ref{eq_maxwell}), we get
\begin{equation}\label{eq_dV_1}
V'(r)=\frac{\alpha_{0}h}{r^{2}}.
\end{equation}
Upon demanding $V'\rightarrow\frac{Q}{r^2}$ and $h\rightarrow1$ as $r\rightarrow\infty$, we can fix the integration constant as $\alpha_{0}=Q$.

Extracting the metric variables $g_{\mu\nu}$ and $q_{\mu\nu}$ from (\ref{eq_metric}) and (\ref{eq_auxi_metric}), and using them in the field equation (\ref{eq_eom2}), we obtain the equations of motion, with $tt$, $rr$, and $\theta\theta$ (or $\varphi\varphi$)  components as

\begin{equation}\label{eq_fieldeq_1}
\frac{C^{2}}{AB}=\frac{1}{fhr^{2}}\left(\lambda r^{4}+\kappa Q^{2}\right),
\end{equation}

\begin{equation}\label{eq_fieldeq_2}
ABC^{2}=\frac{fh}{r^{2}}\left(\lambda r^{4}+\kappa Q^{2}\right),
\end{equation}

 and

\begin{equation}\label{eq_Aux_B}
B=\frac{h}{r^{4}}\left(\lambda r^{4}-\kappa Q^{2}\right).
\end{equation}
Eliminating $B$ from the above equations leads to
\begin{equation}\label{eq_Aux_A}
A=fr^{4}\left(\lambda r^{4}-\kappa Q^{2}\right)^{-1},
\end{equation}
and
\begin{equation}\label{eq_Aux_C}
C=\frac{1}{r}\sqrt{\lambda r^{4}+\kappa Q^{2}}.
\end{equation}

The Ricci tensor $R_{\mu\nu}(\Gamma)$ depends on the affine connection $\Gamma^{\mu}_{\alpha\beta}$, which is defined in terms of the auxiliary metric $q_{\mu\nu}$ expressed by (\ref{eq_affine}). Thus using (\ref{eq_auxi_metric}), all components of the field equation (\ref{eq_eom1}) can be obtained, with the $tt$, $rr$, and $\theta\theta$ (or $\varphi\varphi$) components as

\begin{equation}\label{eq_eom1_tt}
3\frac{A'}{A}\frac{B'}{B}+2\frac{A'}{A}\frac{C'}{C}+4\frac{B'}{B}\frac{C'}{C}+\frac{A''}{A}+2\frac{B''}{B}=\frac{2}{\kappa A}\left(\frac{r^{4}}{\lambda r^{4}-\kappa Q^{2}}-1\right),
\end{equation}

\begin{equation}\label{eq_eom1_rr}
3\frac{A'}{A}\frac{B'}{B}+2\frac{A'}{A}\frac{C'}{C}+\frac{A''}{A}+2\frac{B''}{B}+4\frac{C''}{C}=\frac{2}{\kappa A}\left(\frac{r^{4}}{\lambda r^{4}-\kappa Q^{2}}-1\right),
\end{equation}

and

\begin{equation}\label{eq_eom1_thth}
\frac{C''}{C}+\frac{A'}{A}\frac{C'}{C}+\frac{B'}{B}\frac{C'}{C}+\left(\frac{C'}{C}\right)^{2}-\frac{1}{AC^{2}}=\frac{1}{\kappa A}\left(\frac{r^{4}}{\lambda r^{4}+\kappa Q^{2}}-1\right).
\end{equation}

From equations (\ref{eq_eom1_tt}) and (\ref{eq_eom1_rr}) we find

\begin{equation}\label{eq_ddc_db_rel}
 \frac{C''}{C'}=\frac{B'}{B}.
\end{equation}

Integrating this equation, we get the relation $B=\alpha_{1}C'$, which leads to $h=\frac{\alpha_{1}r^{2}}{\sqrt{\lambda r^{4}+\kappa Q^{2}}}$, upon using (\ref{eq_Aux_B}) and (\ref{eq_Aux_C}). Demanding $h\rightarrow1$ as $r\rightarrow\infty$, we fix the integration constant $\alpha_{1}=\sqrt{\lambda}$, giving
\begin{equation}\label{eq_h}
h=\frac{\sqrt{\lambda}r^{2}}{\sqrt{\lambda r^{4}+\kappa Q^{2}}},
\end{equation}
and substituting (\ref{eq_h}) in (\ref{eq_dV_1}) with $\alpha_0=Q$, we get
\begin{equation}\label{eq_dV}
V'(r)=\frac{\sqrt{\lambda}Q}{\sqrt{\lambda r^{4}+\kappa Q^{2}}}.
\end{equation}

We obtain the differential equation for the metric coefficient $f(r)$ by employing (\ref{eq_Aux_B}), (\ref{eq_Aux_A}),  (\ref{eq_Aux_C}),  and (\ref{eq_ddc_db_rel}) in (\ref{eq_eom1_thth}), leading to
\begin{equation}\label{eq_df}
f'(r)+\frac{\left\{ \kappa^{2}Q^{4}+6\lambda\kappa Q^{2}r^{4}+\lambda^{2}r^{8}\right\} }{r\left\{ \lambda^{2}r^{8}-\kappa^{2}Q^{4}\right\} }f(r)+\frac{1}{r^{3}}\left(Q^{2}-r^{2}\right)+\Lambda r=0.
\end{equation}

\section{Nature of EiBI-Maxwell black hole spacetime: in-depth analyses}\label{sec_asymp}

In this section, we shall discuss the asymptotic behaviours of the metric coefficients and the Kretschmann scalar near a few points of interest, such as, at long distances, near the center, in intermediate region and near the horizon. We shall also discuss the conditions for the existence of an event horizon.

Using (\ref{eq_Aux_C}) and  (\ref{eq_ddc_db_rel}) in equation (\ref{eq_eom1_thth}) and rearranging, we obtain
\begin{equation}
\left(ACC'^{2}\right)^{\prime}=C'+\frac{C'}{\kappa}\left(r^{2}-C^{2}\right).
\end{equation}
This equation immediately leads to the integral form
\begin{equation}\label{eq_integral_A}
A=\frac{1}{C'^{2}}+\frac{1}{CC'^{2}}\int\frac{C'}{\kappa}\left(r^{2}-C^{2}\right)dr+\alpha_{3},
\end{equation}
where $\alpha_3$ is the integration constant. Substituting for $A(r)$ and $C(r)$ from (\ref{eq_Aux_A}) and (\ref{eq_Aux_C}), we get

\begin{equation}\label{eq_f_int_form}
f(r)=\frac{\lambda r^{4}+\kappa Q^{2}}{\lambda r^{4}-\kappa Q^{2}}+\frac{r\sqrt{\lambda r^{4}+\kappa Q^{2}}}{\lambda r^{4}-\kappa Q^{2}}\left[\int\frac{\left\{ Q^{2}+\Lambda r^{4}\right\} \left\{ \kappa Q^{2}-\lambda r^{4}\right\} }{r^{4}\sqrt{\lambda r^{4}+\kappa Q^{2}}}dr-2\sqrt{\lambda}M\right],
\end{equation}
where $-2\sqrt{\lambda}M$ is the integration constant $\alpha_3$ fixed by taking the asymptotic limit $r\rightarrow\infty$ and demanding that $f(r)$ goes over to the de Sitter form $f(r)=1-\frac{2M}{r}+\frac{Q^2}{r^2}-\frac{1}{3}\Lambda r^2$.

In addition, the metric (\ref{eq_metric}) gives the Kretschmann scalar $\mathcal{K}=R^{\alpha\beta\gamma\delta}R_{\alpha\beta\gamma\delta}$ in the form
\begin{equation}\label{eq_krechmann}
 \begin{aligned}\mathcal{K}(r)= & \frac{2f'(r)^{2}}{r^{2}}+\frac{4\left[f(r)-1\right]^{2}}{r^{4}}+\frac{2\left[h(r)f'(r)+2f(r)h'(r)\right]^{2}}{r^{2}h(r)^{2}}\\
 & +\frac{\left[h(r)f''(r)+3f'(r)h'(r)+2f(r)h''(r)\right]^{2}}{h(r)^{2}}.
\end{aligned}
\end{equation}

We shall consider the cosmological constant $\Lambda=0$ in the rest of the discussion, so that $\lambda=1$. Consequently, $f(r)$ would approach Reissner-Nordstr\"om spacetime as $r\rightarrow\infty$.

\subsection{Long distance behaviour}

In order to find the long distance behaviour of the metric coefficient $f(r)$, we expand (\ref{eq_f_int_form}) in the asymptotic limit $r\rightarrow\infty$, giving
\begin{equation}
f(r)\approx1+\frac{1}{r}\left\{ -\int\frac{Q^{2}}{r^{2}}dr-2M\right\} .
\end{equation}
with $\Lambda=0$ and $\lambda=1$. This immediately yields the Reissner-Nordstr\"om metric
\begin{equation}\label{eq_NR}
 f(r)=1-\frac{2M}{r}+\frac{Q^2}{r^2},
\end{equation}
noting that $fh^2=f$ since $h\rightarrow1$ as $r\rightarrow\infty$ from equation (\ref{eq_h}).

The above asymptotic form (\ref{eq_NR}) can also be obtained from equation (\ref{eq_df}) which has the asymptotic form
\begin{equation}\label{eq_df_asym}
 f'(r)+\frac{1}{r}f(r)-\frac{1}{r}+\frac{Q^2}{r^3}\approx0,
\end{equation}
whose solution is immediately obtained as (\ref{eq_NR}) with integration constant $-2M$, by demanding that the spacetime goes over to the Reissner-Nordstr\"om form as $r\rightarrow\infty$. The constant $M$ can be interpreted as the ADM mass of the black hole as measured by an observer at $r\rightarrow\infty$. It is evident that $f(r)$ approaches general relativistic solution.

Upon using (\ref{eq_NR}) and the asymptotic form and $h\approx 1-\frac{\kappa Q^2}{2r^4}$ in (\ref{eq_krechmann}), the Kretschmann scalar is found to be
\begin{equation}\label{eq_krechmann_asym}
 \mathcal{K}=\frac{48M^{2}}{r^{6}}
\end{equation}
in the leading order, that approaches zero as $r\rightarrow\infty$.

\subsection{Behaviour near the center}

The behavior near the centre can be obtained by approximating equation (\ref{eq_f_int_form}) in the limit $r\rightarrow0$, giving
\begin{equation}\label{eq_f_r0}
 f(r)=-1+\frac{Q^{2}}{3r^{2}}+\frac{2Mr}{\sqrt{\kappa Q^{2}}}.
\end{equation}
Thus the metric coefficient $f(r)\rightarrow\infty$ as $r\rightarrow0$.

Moreover, we see from equation (\ref{eq_h}) that $h^2\rightarrow\frac{ r^{4}}{\kappa Q^{2}}$ as $r\rightarrow0$. Consequently,
\begin{equation}\label{eq_fh2_r0_1}
 fh^{2}=\frac{r^{4}}{\kappa Q^{2}}\left(\frac{2Mr}{\sqrt{\kappa Q^{2}}}+\frac{Q^{2}}{3r^{2}}-1\right)
\end{equation}
as $r\rightarrow0$, irrespective of the value of $Q\neq0$.

Upon using (\ref{eq_f_r0}) and $h\approx \frac{r^{2}}{\sqrt{\kappa Q^{2}}}$ in (\ref{eq_krechmann}), the Kretschmann scalar is found to be
\begin{equation}\label{eq_krechmann_near0_Q}
 \mathcal{K}=\frac{52Q^{4}}{9r^{8}}
\end{equation}
in the leading order, which diverges strongly as $r\rightarrow0$.

For the chargeless case $Q=0$, equation (\ref{eq_f_int_form}) gives
\begin{equation}\label{eq_SC}
 f(r)=1-\frac{2M}{r},
\end{equation}
suggesting that $f(r)\rightarrow-\infty$ as $r\rightarrow0$ in the Schwarzchild case. Moreover, we see from equation (\ref{eq_h}) that $h=1$ for $Q=0$. Consequently,
\begin{equation}\label{eq_fh2_near0_2}
 fh^{2}=1-\frac{2M}{r}
\end{equation}
as $r\rightarrow0$ for the chargeless case $Q=0$.

The above solution (\ref{eq_SC}) can also be obtained from equation (\ref{eq_df}), which has the asymptotic behaviour
\begin{equation}
 f^{\prime}(r)+\frac{f(r)}{r}-\frac{1}{r}=0
\end{equation}
near the center, $r\rightarrow0$, immediately leading to the solution (\ref{eq_SC}) with $2M$ as the integration constant.

Upon using (\ref{eq_SC})  and $h=1$ in (\ref{eq_krechmann}), the Kretschmann scalar is found to be
\begin{equation}\label{eq_krechmann_near0}
 \mathcal{K}=\frac{48M^2}{r^6},
\end{equation}
that diverges strongly as $r\rightarrow0$.

\subsection{Intermediate behaviour}

\subsubsection{The case of positive EiBI coupling}

Equation (\ref{eq_f_int_form}) shows that the metric coefficient $f(r)$ diverges when $r^4=\kappa Q^2$ with $\lambda=1$. To find the singular behavior around this point, we substitute $r=a+x$, where $a=\left(\kappa Q^{2}\right)^{1/4}$, and expand about $x=0$. In the leading order, equation (\ref{eq_f_int_form}) takes the form
\begin{equation}\label{eq_f_sing}
f(x)=\frac{C_{-1}}{x}+C_{0}+C_{1}x+C_{2}x^{2}+\ldots,
\end{equation}
where
\begin{equation}\label{eq_f_sing_coef}
\begin{aligned}C_{-1} & =\frac{a}{2}-\frac{M}{\sqrt{2}},\\
C_{0} & =\frac{1}{4}-\frac{M}{2\sqrt{2}a},\\
C_{1} & =-\frac{M}{4\sqrt{2}a^{2}}-\frac{Q^{2}}{2a^{3}}+\frac{5}{8a},\\
C_{2} & =\frac{11Q^{2}}{12a^{4}}-\frac{5}{16a^{2}}+\frac{M}{8\sqrt{2}a^{3}}.
\end{aligned}
\end{equation}
This shows that $f(x)\rightarrow\infty$ upon approaching $x\rightarrow0$, equivalently $r\rightarrow a=\left(\kappa Q^{2}\right)^{1/4}$.

Substituting $r=a+x$ in equation (\ref{eq_h}) with $\lambda=1$, and expanding around $x=0$, we have
\begin{equation}\label{eq_h_sing}
h(x)=\frac{1}{\sqrt{2}}\left(1+\frac{x}{a}-\frac{x^{2}}{a^{2}}+\ldots\right).
\end{equation}

Consequently, the behaviour of the metric coefficient $f(x)h^2(x)$ near $x=0$, or equivalently near  $r=a$, is obtained as
\begin{equation}\label{eq_fh2_sing}
 f(x)h^{2}(x)=\frac{C_{-1}}{2x}+\left(\frac{C_{-1}}{a}+\frac{C_{0}}{2}\right)+\left(-\frac{C_{-1}}{2a^{2}}+\frac{C_{0}}{a}+\frac{C_{1}}{2}\right)x+\ldots,
\end{equation}
implying that $f(x)h^2(x)\rightarrow\infty$ as $x\rightarrow0$, or equivalently $r\rightarrow a=\left(\kappa Q^{2}\right)^{1/4}$.

Using (\ref{eq_f_sing}) and (\ref{eq_h_sing}) in (\ref{eq_krechmann}), we obtain the Kretschmann scalar
\begin{equation}\label{eq_krech_sing}
\mathcal{K}(x)=\frac{\left(a-\sqrt{2}M\right)^{2}}{x^{6}}
\end{equation}
in the leading order. This shows that the Kretschmann scalar $\mathcal{K}\rightarrow\infty$ upon approaching $x\rightarrow0$, equivalently $r\rightarrow a=\left(\kappa Q^{2}\right)^{1/4}$.

\subsubsection{The case of negative EiBI coupling}

Equations (\ref{eq_h}) and (\ref{eq_f_int_form}) show that the metric coefficients $f(r)$ and $h(r)$ diverge when $r^4=|\kappa| Q^2$ with $\lambda=1$. To find the singular behavior around this point, we substitute $r=b+x$, where $b=\left(|\kappa| Q^{2}\right)^{1/4}$, and expand about $x=0$. In the leading order, equation (\ref{eq_f_int_form}) takes the form
\begin{equation}\label{eq_f_sing_kn}
f(x)=-\alpha_{1/2}x^{1/2}+\alpha_{1}x+\alpha_{3/2}x^{3/2}+\alpha_{2}x^{2}+\ldots,
\end{equation}
where
\begin{equation}\label{eq_f_int_alpha_coef}
 \begin{aligned}\alpha_{1/2} & =\frac{2M}{b^{3/2}},\\
\alpha_{1} & =\frac{2}{b}-\frac{2Q^{2}}{b^{3}},\\
\alpha_{3/2} & =\frac{M}{2b^{5/2}},\\
\alpha_{2} & =\frac{7Q^{2}}{3b^{4}}-\frac{1}{b^{2}},
\end{aligned}
\end{equation}
which shows that $f(x)\rightarrow0$ upon approaching $x\rightarrow0$, equivalently $r\rightarrow b=\left(|\kappa| Q^{2}\right)^{1/4}$.

Substituting $r=b+x$ in equation (\ref{eq_h}) with $\lambda=1$, and expanding around $x=0$, we have
\begin{equation}\label{eq_h_sing_kn}
h(x)=\frac{\beta_{-1/2}}{x^{1/2}}+\beta_{1/2}x^{1/2}+\beta_{3/2}x^{3/2}+\ldots,
\end{equation}
where
\begin{equation}
 \begin{aligned}\beta_{-1/2} & =\frac{b^{1/2}}{2},\\
\beta_{1/2} & =\frac{5}{8b^{1/2}},\\
\beta_{3/2} & =-\frac{5}{64b^{3/2}},
\end{aligned}
\end{equation}
so that $h(x)\rightarrow\infty$ upon approaching $x\rightarrow0$, or $r\rightarrow b=\left(|\kappa| Q^{2}\right)^{1/4}$.

Consequently, the behaviour of the metric coefficient $f(x)h^2(x)$ near $x=0$, or equivalently near  $r=b$, is obtained as
\begin{equation}\label{eq_fh2_sing_kn}
f(x)h^{2}(x)=-\frac{\gamma_{-1/2}}{x^{1/2}}+\gamma_{0}+\gamma_{1/2}x^{1/2}+\gamma_{1}x+\gamma_{3/2}x^{3/2}+\ldots,
\end{equation}
where
\begin{equation}
 \begin{aligned}\gamma_{-1/2} & =\frac{M}{2b^{1/2}},\\
\gamma_{0} & =\frac{b^{2}-Q^{2}}{2b^{2}},\\
\gamma_{1/2} & =-\frac{9M}{8b^{3/2}},\\
\gamma_{1} & =\frac{1}{b}-\frac{2Q^{2}}{3b^{3}},\\
\gamma_{3/2} & =-\frac{5M}{16b^{5/2}},
\end{aligned}
\end{equation}
implying that $f(x)h^2(x)\rightarrow\infty$ as $x\rightarrow0$, or equivalently $r\rightarrow b=\left(|\kappa| Q^{2}\right)^{1/4}$.

Using (\ref{eq_f_sing_kn}) and (\ref{eq_h_sing_kn}) in (\ref{eq_krechmann}), the Kretschmann scalar is found to be
\begin{equation}\label{eq_krech_sing_kn}
\mathcal{K}(x)=\frac{\left(b-\sqrt{2}M\right)^{2}}{x^{6}}
\end{equation}
in the leading order. This shows that the Kretschmann scalar $\mathcal{K}\rightarrow\infty$ upon approaching $x\rightarrow0$, or equivalently $r\rightarrow b=\left(|\kappa| Q^{2}\right)^{1/4}$.

In summary, for the case $\kappa>0$, $f(r)$ and $f(r)h^2(r)$ diverge at $r=\left(\kappa Q^{2}\right)^{1/4}$ with $h(r)$ approaching a constant. On the other hand, for the case $\kappa<0$, $h(r)$ and $f(r)h^2(r)$ diverge at $r=\left(|\kappa| Q^{2}\right)^{1/4}$ with $f(r)$ approaching zero. However, the Kretschmann scalar $\mathcal{K}(r)$ diverges strongly in both cases, $\kappa>0$ and $\kappa<0$.

\subsection{Near-horizon behaviour}

We can find the behaviour around the event horizon at $r=R$, by writing $r=R+x$ in equation (\ref{eq_df}) and expanding about $x=0$. This yields
\begin{equation}\label{eq_df_horizon}
f^{\prime}(x)+\left(a_{0}+a_{1}x+a_{2}x^{2}\right)f(x)+\left(b_{0}+b_{1}x+b_{2}x^{2}\right)\approx0,
\end{equation}
where
\begin{equation}\label{eq_ab_coef}
\begin{aligned}a_{0} & =\frac{\kappa^{2}Q^{4}+6\kappa Q^{2}R^{4}+R^{8}}{R\left(R^{8}-\kappa^{2}Q^{4}\right)},\\
a_{1} & =\frac{\kappa^{4}Q^{8}-18\kappa^{3}Q^{6}R^{4}-16\kappa^{2}Q^{4}R^{8}-30\kappa Q^{2}R^{12}-R^{16}}{R^{2}\left(R^{8}-\kappa^{2}Q^{4}\right)^{2}},\\
a_{2} & =\frac{1}{R^{3}\left(R^{8}-\kappa^{2}Q^{4}\right)^{3}}\left[\kappa^{6}Q^{12}+18\kappa^{5}Q^{10}R^{4}+39\kappa^{4}Q^{8}R^{8}+276\kappa^{3}Q^{6}R^{12}\right.\\
 & \left.\ \ \ \ \ \ \ \ \ \ \ \ \ \ \ \ \ \ \ \ \ \ \ \ \ \ \ \ \ \ +87\kappa^{2}Q^{4}R^{16}+90\kappa Q^{2}R^{20}+R^{24}\right]\\
b_{0} & =\frac{Q^{2}-R^{2}}{R^{3}},\\
b_{1} & =\frac{R^{2}-3Q^{2}}{R^{4}},\\
b_{2} & =\frac{6Q^{2}-R^{2}}{R^{5}}.
\end{aligned}
\end{equation}
The solution of equation (\ref{eq_df_horizon}) is given by
\begin{equation}\label{eq_f_horizon_wc}
f(x)=-\left(\frac{b_{0}}{a_{0}}-\frac{b_{1}}{a_{0}^{2}}+\frac{2b_{2}}{a_{0}^{3}}\right)-\left(\frac{b_{1}}{a_{0}}-\frac{2b_{2}}{a_{0}^{2}}\right)x-\frac{b_{2}}{a_{0}}x^{2}+C\left(1-a_{0}x+\frac{a^{2}_{0}}{2}x^{2}\right)+\ldots,
\end{equation}
where $C$ is the integration constant.

It is well-known that EiBI gravity reduces to general relativity in the absence of matter. This fact is obvious from equation (\ref{eq_df}) by setting $Q=0$, leading to
\begin{equation}\label{eq_de_gr}
 f^{\prime}(r)+\frac{1}{r}f(r)-\frac{1}{r}=0,
\end{equation}
giving the solution
\begin{equation}\label{eq_f_gr}
 f(r)=1-\frac{2M}{r},
\end{equation}
where $-2M$ is the integration constant.

The above property should be recovered when we set $Q=0$ in our solution given by equation (\ref{eq_f_horizon_wc}).

Setting $Q\rightarrow0$ in equation (\ref{eq_ab_coef}), we have $a_0\rightarrow1/R$, , $a_2\rightarrow1/R^3$, $b_0\rightarrow-1/R$, $b_1\rightarrow1/R^2$, and $b_2\rightarrow-1/R^3$  so that the solution (\ref{eq_f_horizon_wc}) reduces to
\begin{equation}\label{eq_f_horizon_wc_Q0}
 f(x)=\left(4+C\right)-\left(3+C\right)\frac{x}{R}+\frac{x^{2}}{R^{2}}+\ldots.
\end{equation}

Using $r=R+x$ in the Schwarzchild solution (\ref{eq_f_gr}) and  expanding about $x=0$, we get
\begin{equation}\label{eq_f_gr_hor_exp}
f(x)=1-\frac{2M}{R}+\frac{2M}{R^{2}}x+\frac{2M}{R^{3}}x^{2}+\dots.
\end{equation}

Demanding that EiBI black hole reduces to Schwarzchild black hole for $Q=0$, we can therefore compare (\ref{eq_f_horizon_wc_Q0}) with (\ref{eq_f_gr_hor_exp}) and obtain the integration constant as
\begin{equation}\label{eq_C_coe}
 C=-3-\frac{2M}{R}.
\end{equation}

Using this value of the integration constant in (\ref{eq_f_horizon_wc}), we obtain
\begin{equation}\label{eq_f_horizon}
f(x)=c_{0}+c_{1}x+c_{2}x^2+\ldots
\end{equation}
where
\begin{equation}\label{eq_c_coef}
\begin{aligned}\begin{aligned}c_{0} & =-3-\frac{2M}{R}-\frac{b_{0}}{a_{0}}+\frac{b_{1}}{a_{0}^{2}}-\frac{2b_{2}}{a_{0}^{3}},\\
c_{1} & =a_{0}\left(3+\frac{2M}{R}-\frac{b_{1}}{a_{0}^{2}}+\frac{2b_{2}}{a_{0}^{3}}\right),\\
c_{2} & =-\frac{a_{0}^{2}}{2}\left(3+\frac{2M}{R}+\frac{4b_{2}}{a_{0}^{3}}\right).
\end{aligned}
\end{aligned}
        \end{equation}

Substituting $r=R+x$ in equation (\ref{eq_h}) with $\lambda=1$, and expanding around $x=0$, we have
\begin{equation}\label{eq_h_horizon}
 h(x)=k_{0}+k_{1}x+k_{2}x^{2}+\ldots,
\end{equation}
where
\begin{equation}\label{eq_k_coef}
 \begin{aligned}k_{0} & =\frac{R^{2}}{\sqrt{a^{4}+R^{4}}},\\
k_{1} & =\frac{2a^{4}R}{\left(a^{4}+R^{4}\right)^{3/2}},\\
k_{2} & =\frac{\left(a^{8}-5a^{4}R^{4}\right)}{\left(a^{4}+R^{4}\right)^{5/2}}.
\end{aligned}
\end{equation}
Thus, the metric potential $f(r)h^2(r)\rightarrow0$ as $r\rightarrow R$ (or $x\rightarrow0$).

Using (\ref{eq_f_horizon}) and (\ref{eq_h_horizon}) in (\ref{eq_krechmann}), we have the Kretschmann scalar
\begin{equation}\label{eq_krech_hor}
 \mathcal{K}(x)=\mathcal{K}_{0}+\mathcal{K}_{1}x+\mathcal{K}_{2}x^2+\ldots,
\end{equation}
where
\begin{equation}\label{eq_krech_coe_hor}
 \begin{aligned}\mathcal{K}_{0} & =\frac{\left(2c_{2}k_{0}+3c_{1}k_{1}\right){}^{2}}{k_{0}^{2}}+\frac{4c_{1}^{2}}{R^{2}}+\frac{4}{R^{4}},\\
\mathcal{K}_{1} & =\frac{2\left(2c_{2}k_{0}+3c_{1}k_{1}\right)\left(6c_{2}k_{0}k_{1}+10c_{1}k_{0}k_{2}-3c_{1}k_{1}^{2}\right)}{k_{0}^{3}}\\
 & +\frac{4\left(2c_{2}c_{1}k_{0}R+c_{1}^{2}(-k_{0})+2c_{1}^{2}k_{1}R\right)}{k_{0}R^{3}}-\frac{8\left(c_{1}R+2\right)}{R^{5}}+\frac{4\left(2c_{1}c_{2}R-c_{1}^{2}\right)}{R^{3}},\\
\mathcal{K}_{2} & =\frac{1}{k_{0}^{4}R^{6}}\left[16c_{2}^{2}k_{0}^{4}R^{4}-32c_{1}c_{2}k_{0}^{4}R^{3}+16c_{1}^{2}k_{0}^{4}R^{2}-8c_{2}k_{0}^{4}R^{2}\right.\\
 & +32c_{1}k_{0}^{4}R+24c_{1}c_{2}k_{0}^{3}k_{1}R^{4}-16c_{1}^{2}k_{0}^{3}k_{1}R^{3}+64c_{2}^{2}k_{0}^{3}k_{2}R^{6}\\
 & +16c_{1}^{2}k_{0}^{3}k_{2}R^{4}+12c_{2}^{2}k_{0}^{2}k_{1}^{2}R^{6}+164c_{1}c_{2}k_{0}^{2}k_{1}k_{2}R^{6}+100c_{1}^{2}k_{0}^{2}k_{2}^{2}R^{6}\\
 & -60c_{1}c_{2}k_{0}k_{1}^{3}R^{6}-138c_{1}^{2}k_{0}k_{1}^{2}k_{2}R^{6}+27c_{1}^{2}k_{1}^{4}R^{6}+40k_{0}^{4}.
\end{aligned}
\end{equation}
It is obvious from (\ref{eq_krech_hor}) and the first line in (\ref{eq_krech_coe_hor}) that the Kretschmann scalar converges to the finite value $\mathcal{K}_0$ at the horizon $r=R$ (or $x=0$).

%==================================================
\begin{figure}[t!] % Adjust placement using [h]ere, [t]op, [b]ottom, [p]age
\centering
\begin{subfigure}{0.49\textwidth} % Adjust width as needed
\includegraphics[width=\textwidth, height=6cm]{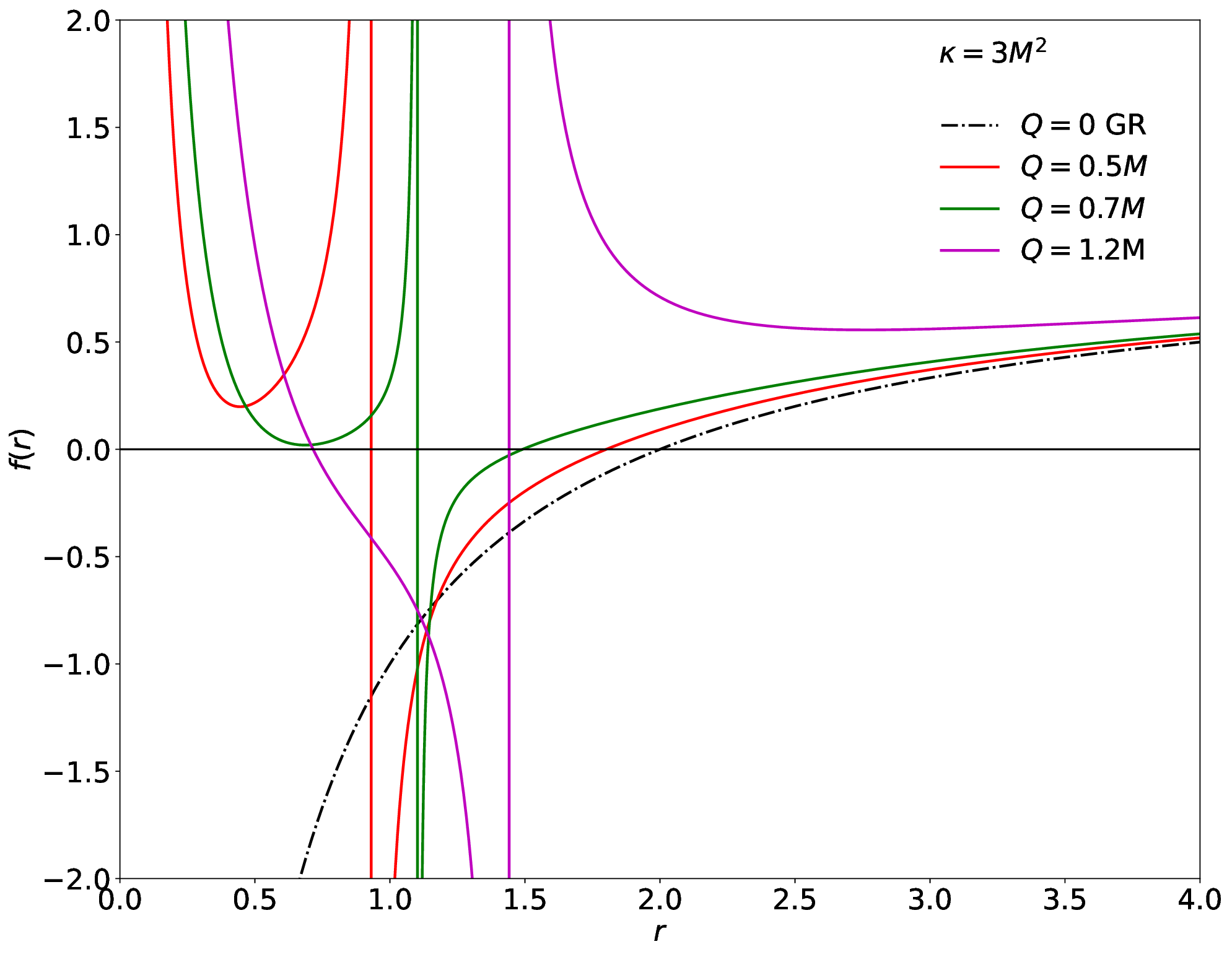} % Replace with your image path
\caption{Radial profiles of $f(r)$ with $\kappa/M^2=3$ for different values of $Q/M$.}
\label{fig:kc_Qv_Pc}
\end{subfigure}
\hfill % Space between subfigures
\begin{subfigure}{0.49\textwidth}
\includegraphics[width=\textwidth, height=6cm]{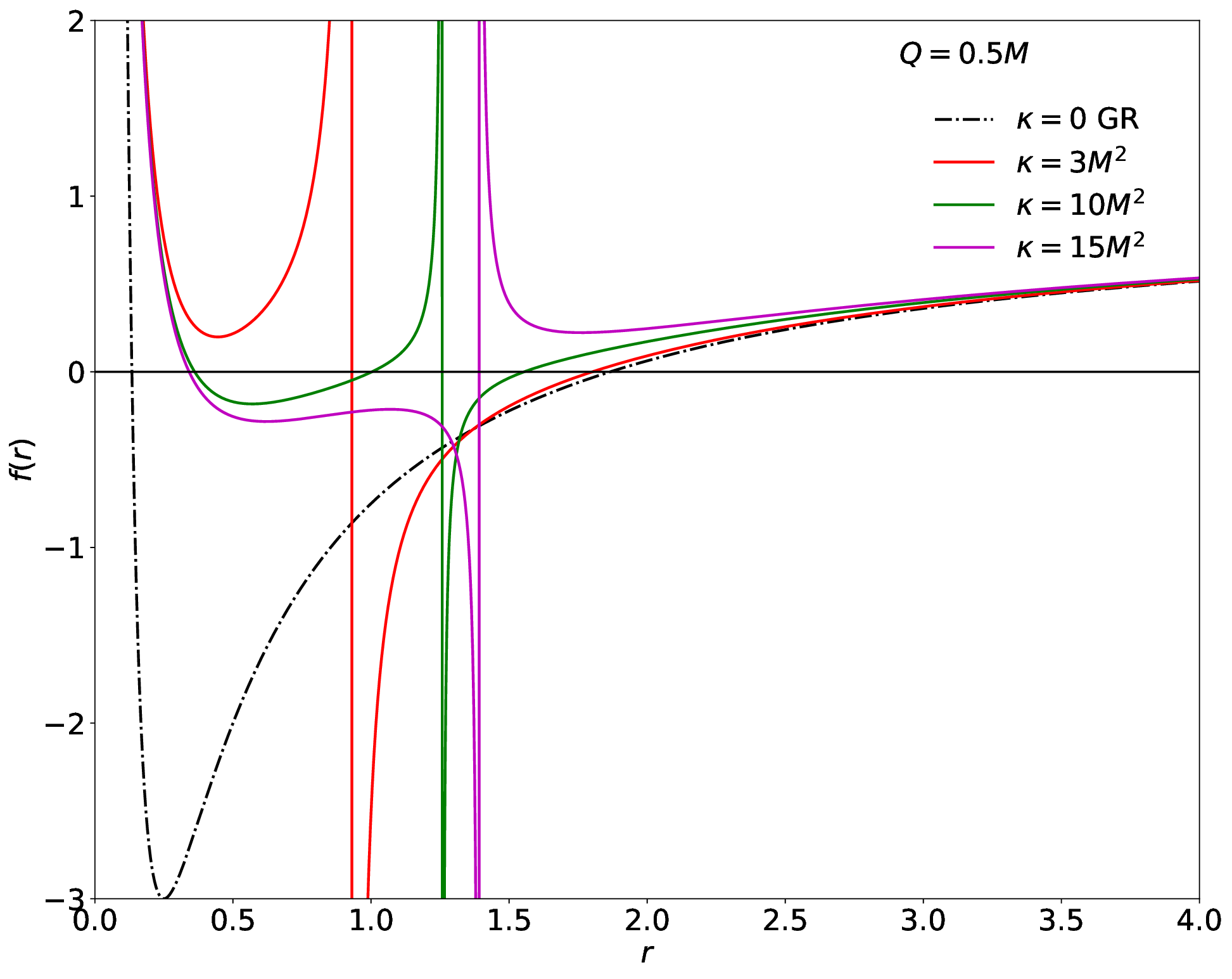}
\caption{Radial profiles of $f(r)$ with $Q/M=0.5$ for different values of $\kappa/M^2>0$. }
\label{fig:kv_Qc_Pc}
\end{subfigure}

\begin{subfigure}{0.49\textwidth} % Adjust width as needed
\includegraphics[width=\textwidth, height=6cm]{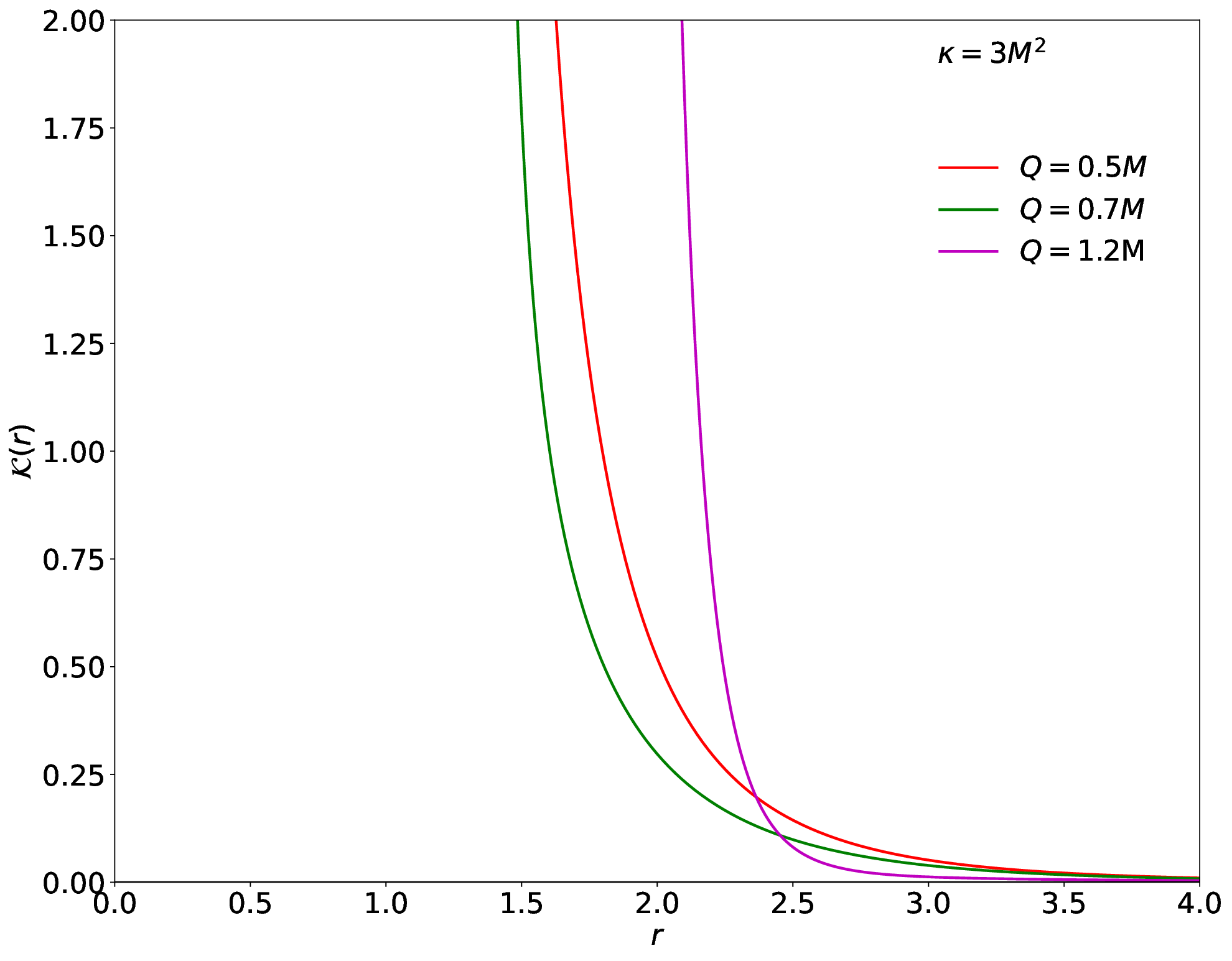} % Replace with your image path
\caption{Radial profiles of $\mathcal{K}(r)$ with $\kappa/M^2=3$ for different values of $Q$.}
\label{fig:krech_Qv_Pc}
\end{subfigure}
\hfill % Space between subfigures
\begin{subfigure}{0.49\textwidth}
\includegraphics[width=\textwidth, height=6cm]{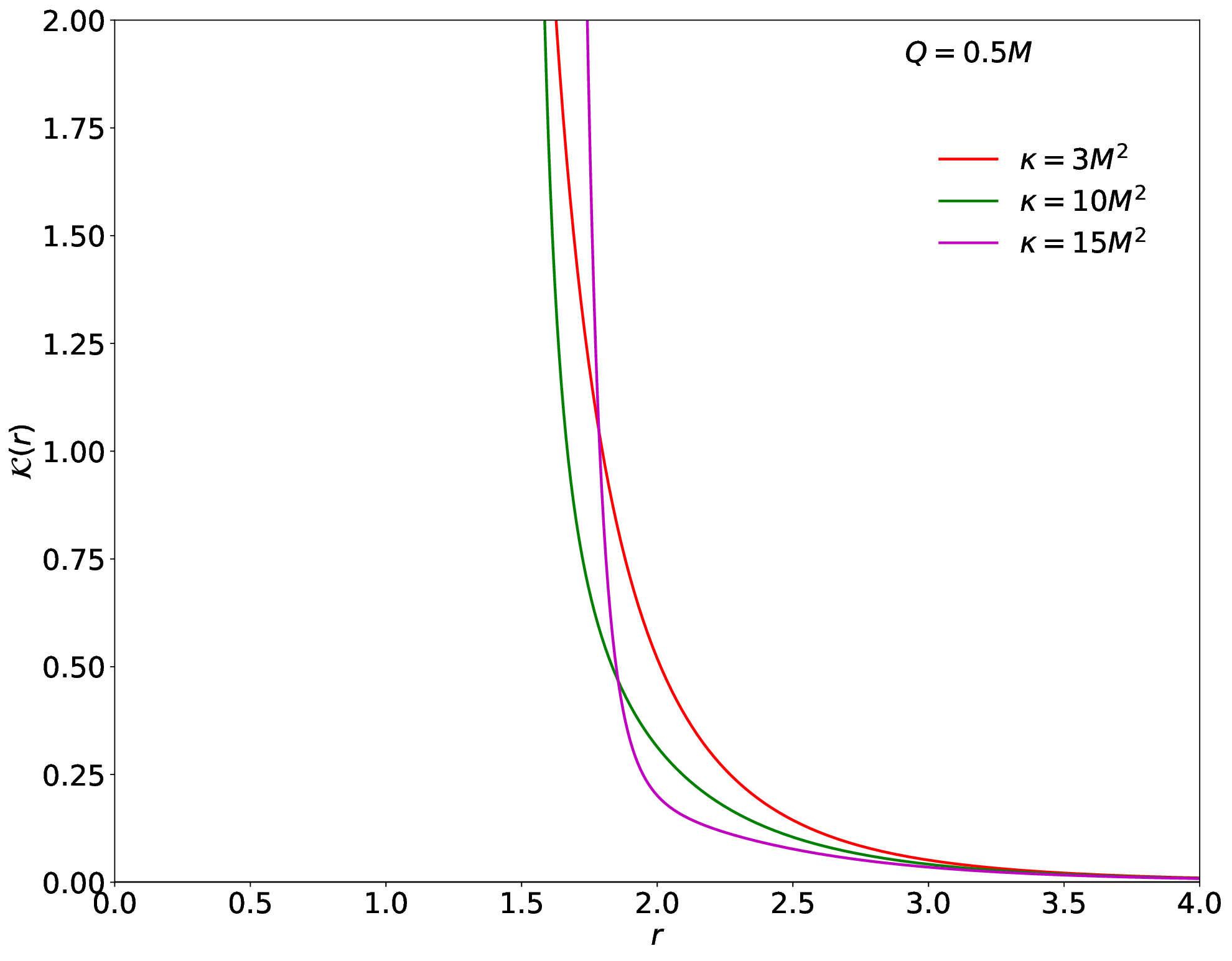}
\caption{Radial profiles of $\mathcal{K}(r)$ with $Q/M=0.5$ for different values of $\kappa/M^2>0$.}
\label{fig:krech_Qc_Pc}
\end{subfigure}
\caption{Radial profiles of the metric potential $f(r)$ and the Kretschmann scalar $\mathcal{K}(r)$ for the EiBI coupling $\kappa>0$.}
\label{fig:kappa_ge_0_P_constant}

\end{figure}

%==================================================
\section{Numerical integration of the field equation}\label{sec_num}

In order to find the nature of spacetime more concretely for charged black holes in EiBI gravity, we shall obtain the radial profiles of the metric potential $f(r)$ and the Kretschmann scalar $\mathcal{K}(r)$ by solving the differential equation (\ref{eq_df}) numerically without any approximation for different values of the parameters $\kappa/M^2$ and $Q/M$.

%==================================================

\begin{figure}[t!]%[htbp] % Adjust placement using [h]ere, [t]op, [b]ottom, [p]age
\centering
\begin{subfigure}{0.49\textwidth} % Adjust width as needed
\includegraphics[width=\textwidth, height=6cm]{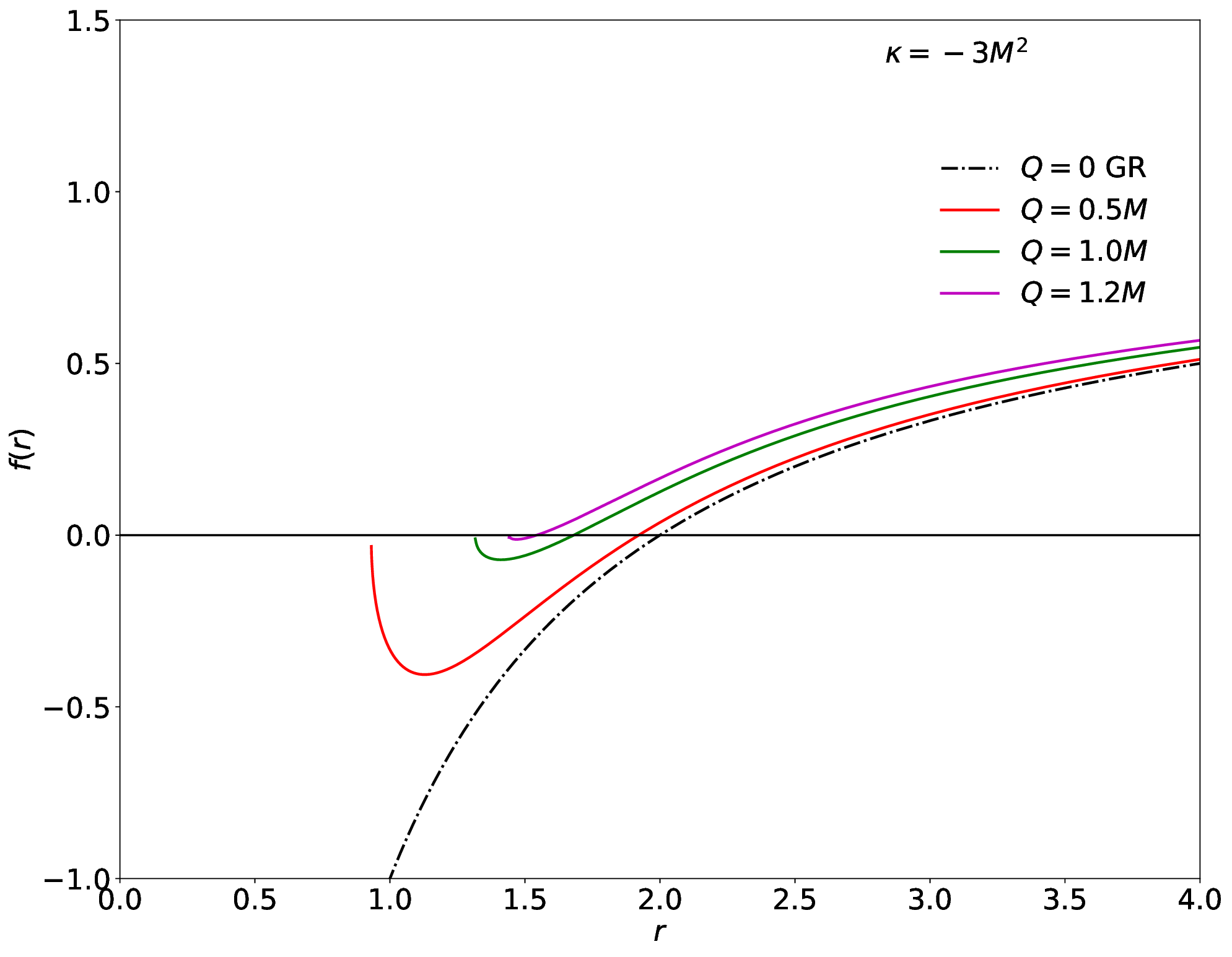} % Replace with your image path
\caption{Radial profiles of $f(r)$ with $\kappa/M^2=-3$ for different values of $Q/M$.}
\label{fig:knc_Qv_Pc}
\end{subfigure}
\hfill % Space between subfigures
\begin{subfigure}{0.49\textwidth}
\includegraphics[width=\textwidth, height=6cm]{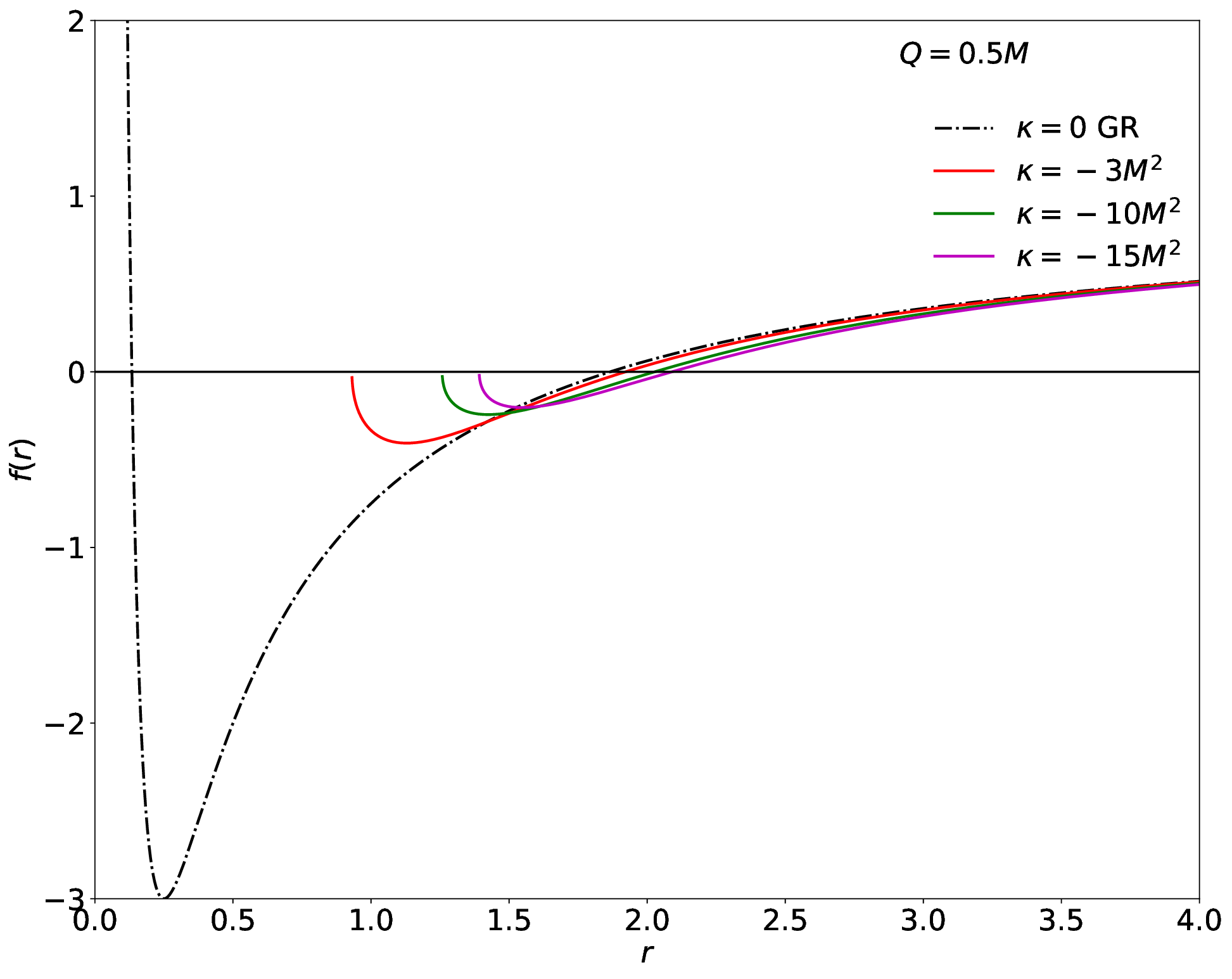}
\caption{Radial profiles of $f(r)$ with $Q/M=0.5$ for different values of $\kappa/M^2<0$.}
\label{fig:knv_Qc_Pc}
\end{subfigure}

\begin{subfigure}{0.49\textwidth} % Adjust width as needed
\includegraphics[width=\textwidth, height=6cm]{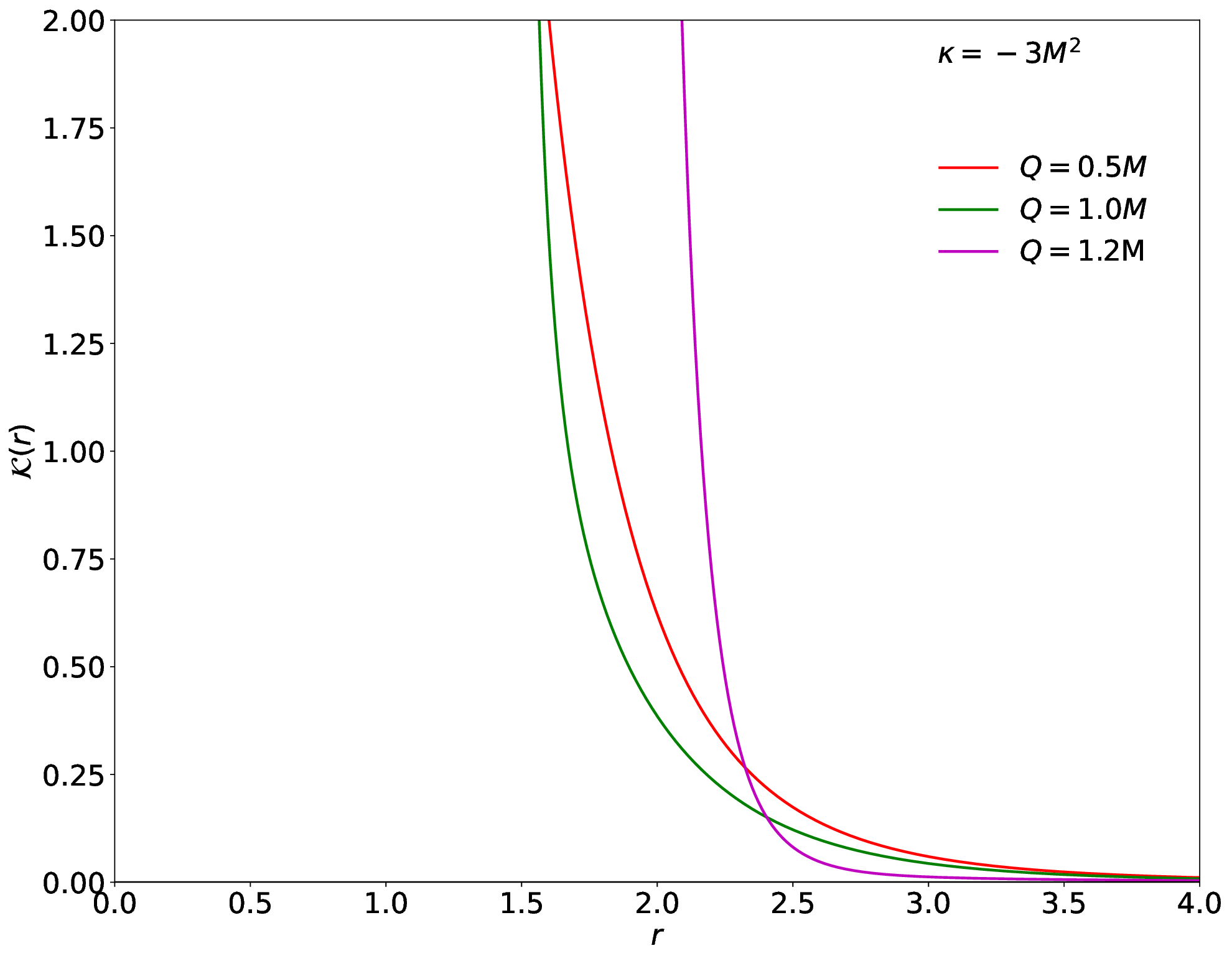} % Replace with your image path
\caption{Radial profiles of $\mathcal{K}(r)$ with $\kappa/M^2=-3$ for different values of $Q/M$.}
\label{fig:krech_knc_Qv_Pc}
\end{subfigure}
\hfill % Space between subfigures
\begin{subfigure}{0.49\textwidth}
\includegraphics[width=\textwidth, height=6cm]{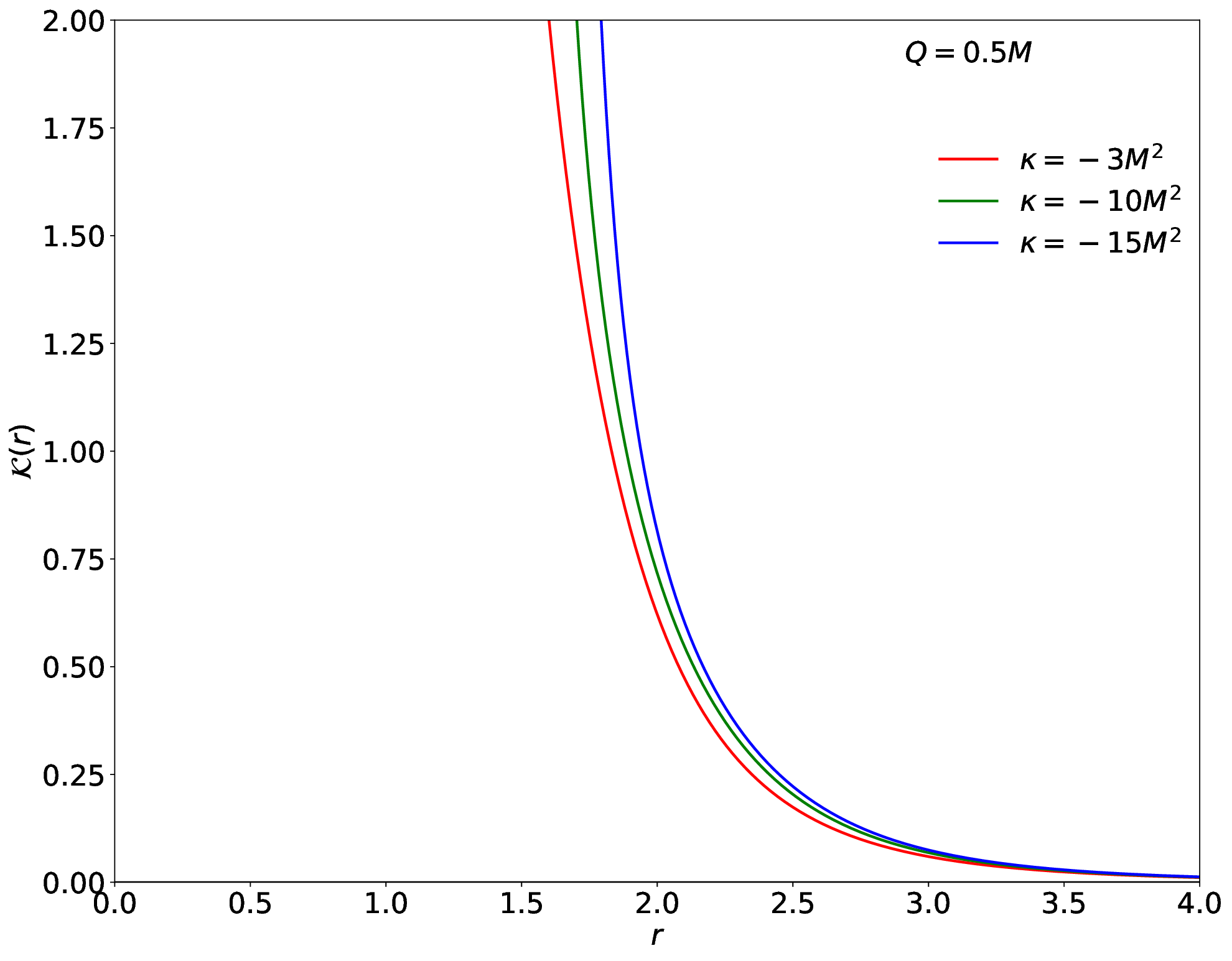}
\caption{Radial profiles of $\mathcal{K}(r)$ with $Q/M=0.5$ for different values of $\kappa/M^2<0$.}
\label{fig:krech_knv_Qc_Pc}
\end{subfigure}
\caption{Radial profiles of the metric potential $f(r)$ and the Kretschmann scalar $\mathcal{K}(r)$ for the EiBI coupling $\kappa<0$.}
\label{fig:kappan_ge_0_P_constant}
\end{figure}

%==================================================
%for different values of $\kappa/M^2$
Figure \ref{fig:kappa_ge_0_P_constant} displays the radial profiles of the metric potential $f(r)$ and the Kretschmann scalar $\mathcal{K}(r)$ for the case $\kappa>0$. Figure \ref{fig:kc_Qv_Pc} shows the behaviour of $f(r)$ for $\kappa/M^2=3$ and different values of electric charge $Q$. We note from equation (\ref{eq_df}) and (\ref{eq_f_sing}) that black holes can exist with an additional singularity at $r=a=\left(\kappa Q^{2}\right)^{1/4}$ in EiBI gravity, which is clearly a departure from general relativity. This singularity is seen to occur for all cases of $\kappa>0$.

Equations (\ref{eq_h}) and (\ref{eq_df}) indicate that the metric coefficients reduce to the Schwarzchild solution for the case $Q=0$. Setting $Q=0$, our numerical results are consistent with this fact as shown by the dotted line in Figure \ref{fig:kc_Qv_Pc}. Thus in the absence of matter, EiBI gravity reduces to general relativity. On the other hand, as the value of $Q$ is increased, the event horizon shifts inwards as evident from figure \ref{fig:kc_Qv_Pc} for the cases $Q/M=0.5$ and $0.7$. For sufficiently large values of $Q$, for example $Q=1.5M$, the event horizon disappears, so that the singularity at $r=a=\left(\kappa Q^{2}\right)^{1/4}$ given by equation (\ref{eq_f_sing}) becomes naked. Thus, unlike the Reissner-Nordstr\"om case, unphysical solutions can exist in EiBI gravity without an event horizon.

Furthermore, Equations (\ref{eq_h}) and (\ref{eq_df}) indicate that the metric coefficients reduce to the Reissner-Nordstr\"om case for $\kappa=0$. Setting $\kappa=0$, our numerical results are consistent with this fact as shown by the dotted line in Figure \ref{fig:kv_Qc_Pc}. Figure \ref{fig:kv_Qc_Pc} also displays radial profiles of $f(r)$ with respect to different values of the coupling parameter $\kappa/M^2>0$ while the electric charge is fixed at $Q=0.5M$. As shown in figure \ref{fig:kv_Qc_Pc}, for $\kappa/M^2=3$ and $10$, the event horizon shifts inwards with increase in the value of the parameter  $\kappa/M^2$. For sufficiently high values of $\kappa/M^2$, for example $\kappa/M^2=16$, an event horizon ceases  to exist and the singularity becomes naked. Such black hole solutions are unphysical.

Figures \ref{fig:krech_Qv_Pc} and \ref{fig:krech_Qc_Pc} show the divergence of the Kretschmann scalar $\mathcal{K}(r)$  at $r=a=\left(\kappa Q^{2}\right)^{1/4}$. As shown by equation (\ref{eq_krech_sing}), this strong divergence is also obvious from these plots.

%==================================================

\begin{figure}[t!]%[htbp] % Adjust placement using [h]ere, [t]op, [b]ottom, [p]age
\centering
\begin{subfigure}{0.49\textwidth} % Adjust width as needed
\includegraphics[width=\textwidth, height=6cm]{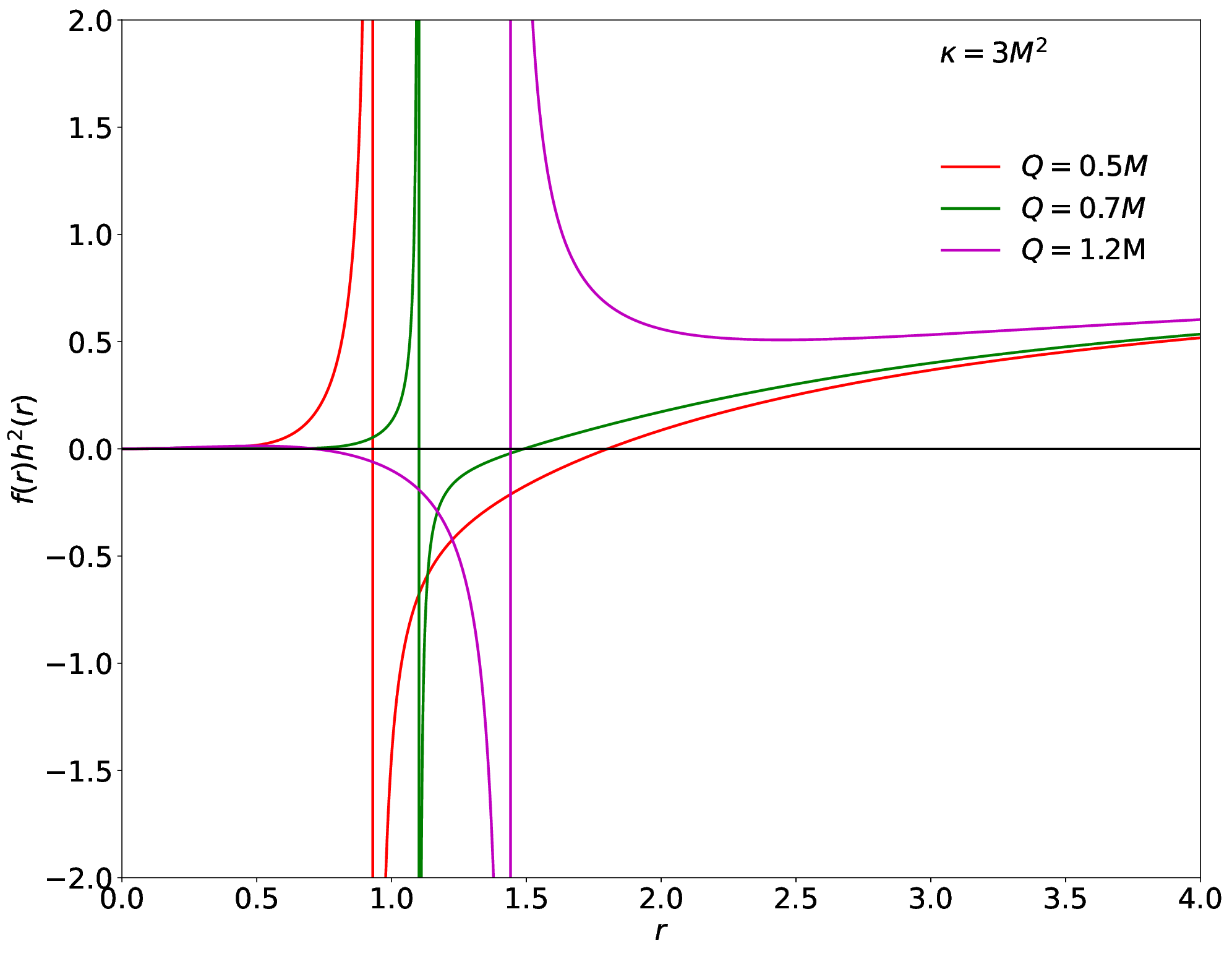} % Replace with your image path
\caption{Radial profiles of $f(r)h^2(r)$ with $\kappa/M^2=3$ for different values of $Q/M$.}
\label{fig:fh2_kpc_Qv_Pc}
\end{subfigure}
\hfill % Space between subfigures
\begin{subfigure}{0.49\textwidth} % Adjust width as needed
\includegraphics[width=\textwidth, height=6cm]{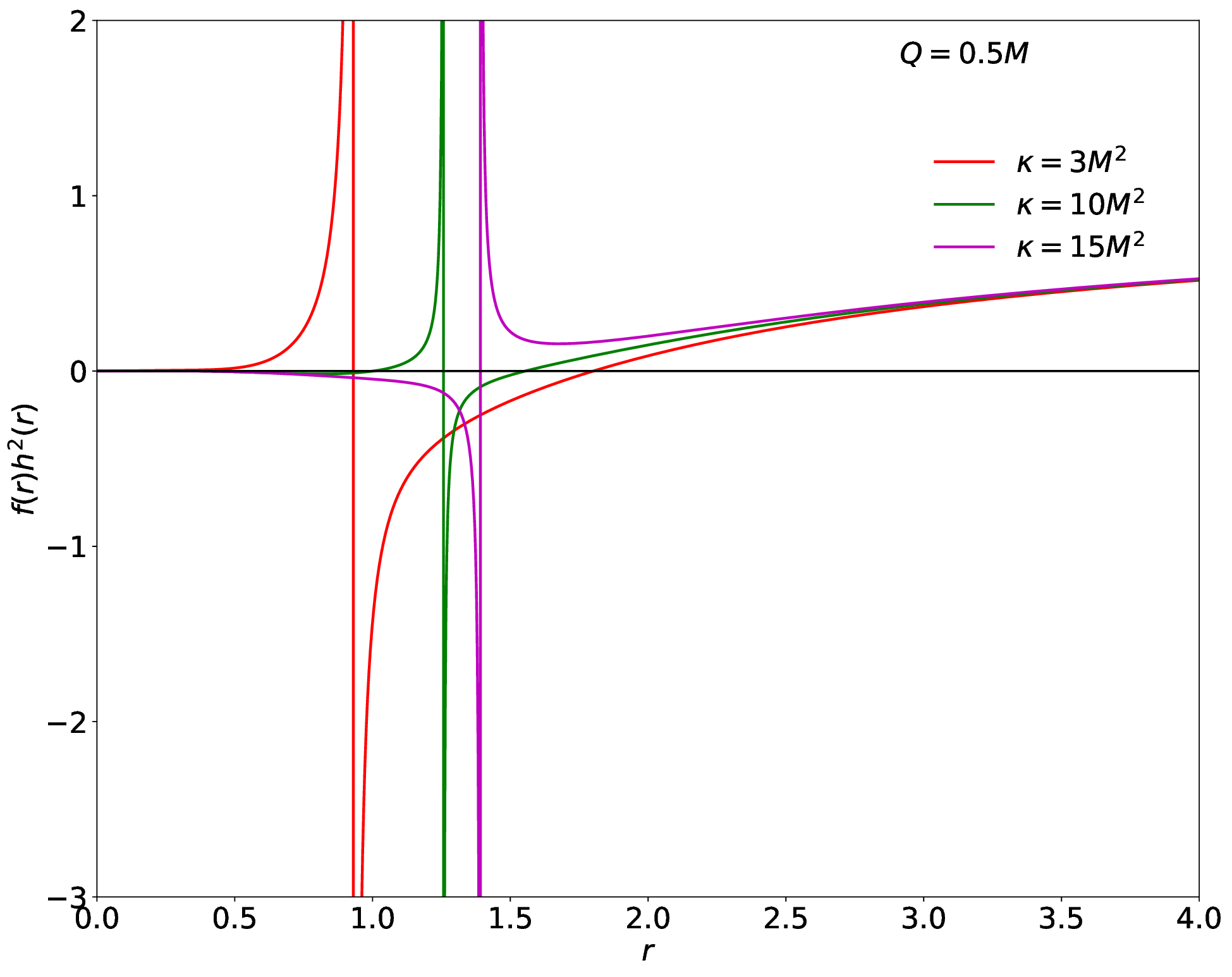} % Replace with your image path
\caption{ Radial profiles of $f(r)h^2(r)$ with $Q/M=0.5$ for different values of $\kappa/M^2>0$.}
\label{fig:fh2_kv_Qc_Pc}
\end{subfigure}

\begin{subfigure}{0.49\textwidth}
\includegraphics[width=\textwidth, height=6cm]{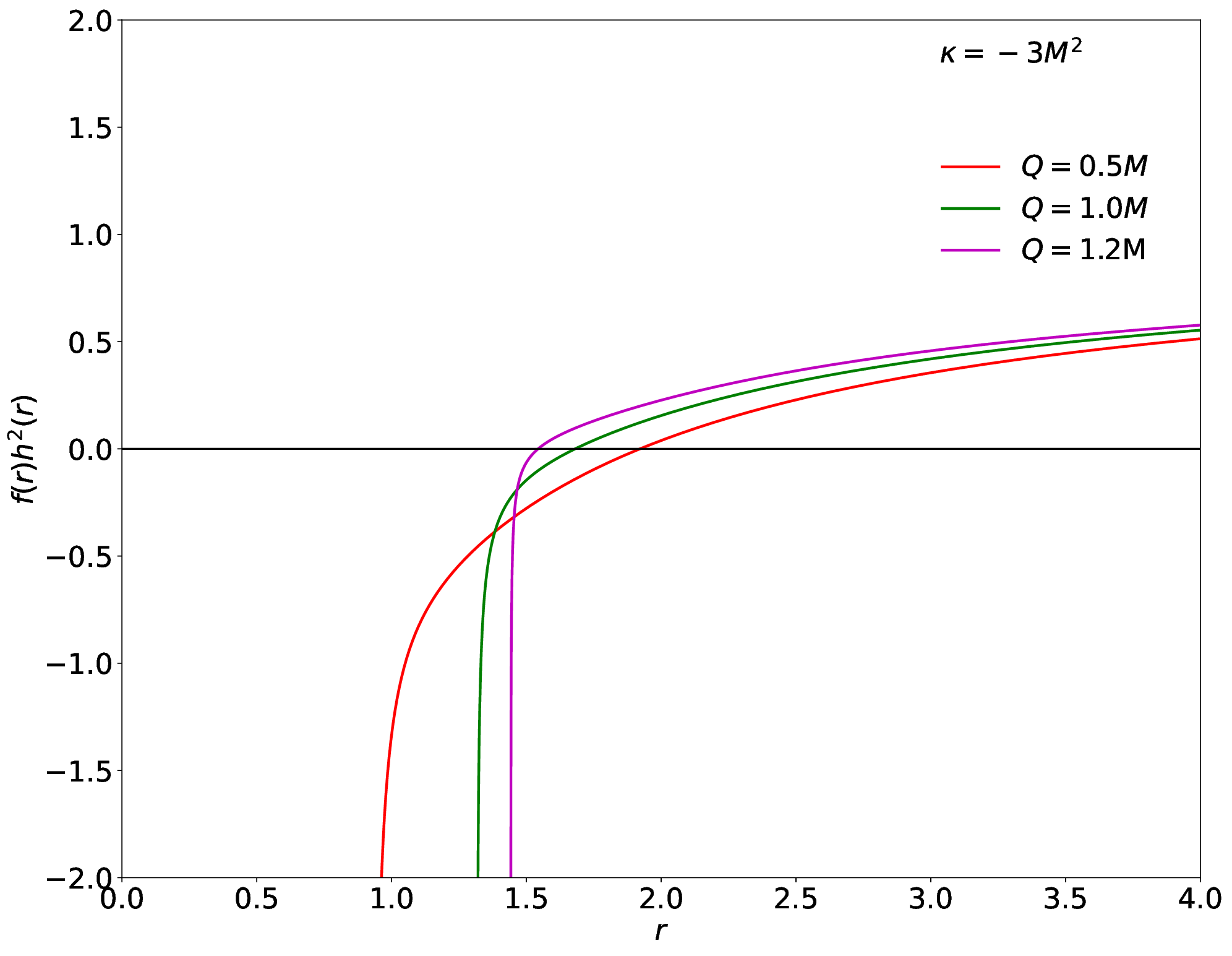}
\caption{Radial profiles of $f(r)h^2(r)$ with $\kappa/M^2=-3$ for different values of $Q/M$.}
\label{fig:fh2_knc_Qv_Pc}
\end{subfigure}
\hfill % Space between subfigures
\begin{subfigure}{0.49\textwidth}
\includegraphics[width=\textwidth, height=6cm]{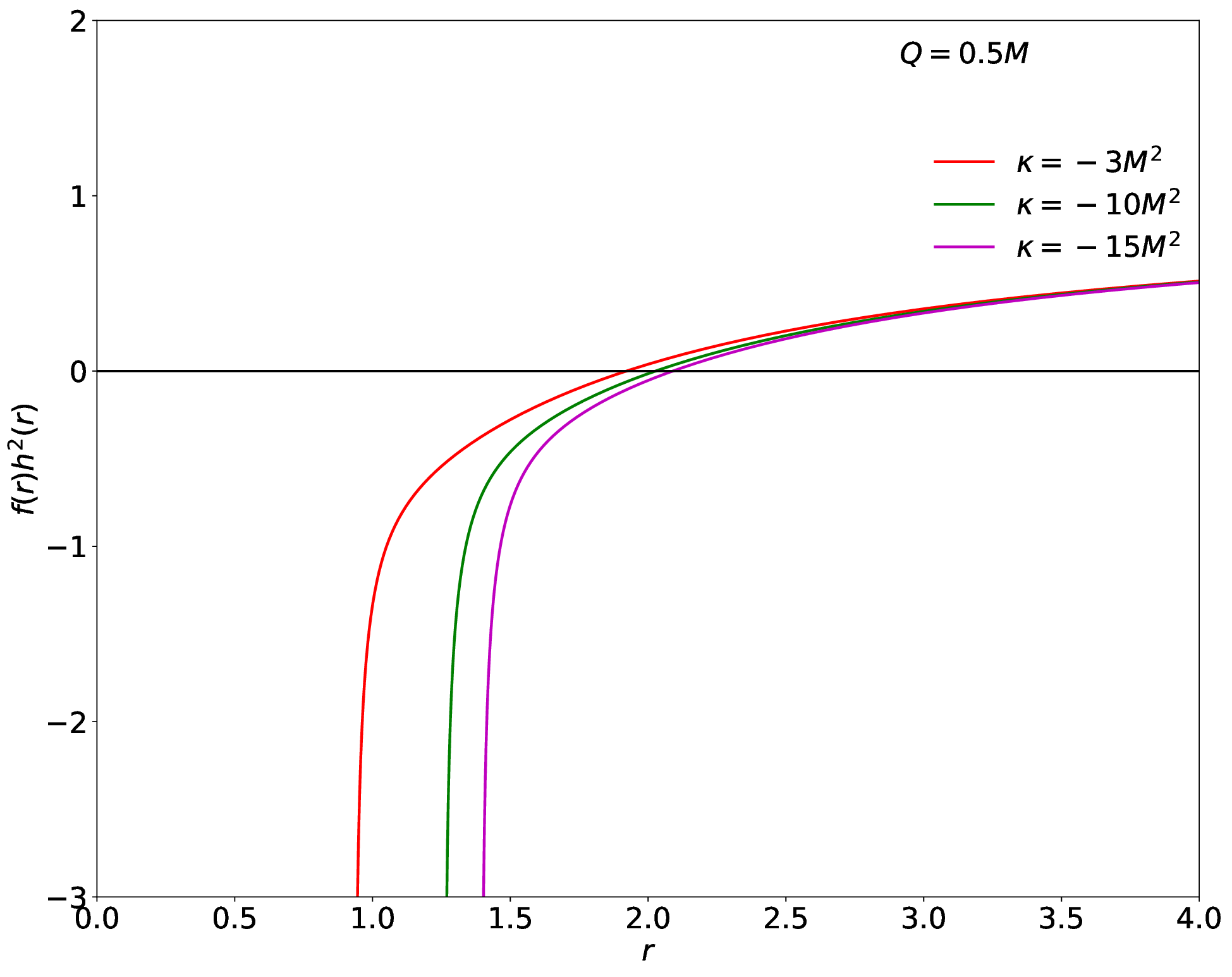}
\caption{Radial profiles of $f(r)h^2(r)$ with $Q/M=0.5$ for different values of $\kappa/M^2<0$.}
\label{fig:fh2_knv_Qc_Pc}
\end{subfigure}
\caption{Radial profiles of the metric potential $f(r)h^2(r)$ for the cases $\kappa>0$ and $\kappa<0$.}
\label{fig:fh2_ge_0_P_constant}
\end{figure}

%==================================================
Figure\ \ref{fig:kappan_ge_0_P_constant} displays radial profiles of the metric potential $f(r)$ and the Kretschmann scalar $\mathcal{K}(r)$ when $\kappa<0$. As seen from equation (\ref{eq_h}), the metric potential $h(r)$ becomes imaginary for $r^4<|\kappa|Q^2$, so that the spacetime becomes unphysical. Consequently, we have not shown the profile for $f(r)$ in the region $0\leq r\leq |\kappa|Q^2$ in figure \ref{fig:kappan_ge_0_P_constant}. Figure\ \ref{fig:knc_Qv_Pc} shows the profiles of $f(r)$ with different values of the electric charge $Q$ while $\kappa=-3M^2$ is fixed. The dotted line shows radial profile of $f(r)$ for $Q=0$, which is the same as the Schwarzchild solution as also obvious from equations (\ref{eq_h}) and (\ref{eq_df}) with $\lambda=1$. We further note that, upon increasing the electric charge $Q$, the horizon shifts inward, as shown for the cases $Q/M=0.5$ to $1.2$ in figure \ref{fig:knc_Qv_Pc}. Consequently, event horizon can exist for black holes carrying electric charge $Q>M$ in EiBI gravity, which is a clear departure from the Reissner-Nordstr\"om solution. However, for sufficiently high values of the charge $Q$, for example $Q/M=1.5$, the horizon disappears and the singularity becomes naked.

Moreover,  Figure \ref{fig:knv_Qc_Pc} displays the profiles of $f(r)$ for different values of $\kappa<0$ while the electric charge $Q=0.5M$ is fixed. The general relativistic case is shown by the dotted line with $\kappa=0$ which is equivalent to the Reissner-Nordstr\"om solution. As $|\kappa|$ is increased, for example for $\kappa/M^2=-0.3$,$-10.$,$-15$, we see that the horizon moves outward.

As shown by equation (\ref{eq_krech_sing_kn}), the Kretschmann scalar diverges strongly at $r=\left(|\kappa|Q^{2}\right)^{1/4}$. This behaviour is displayed in figures \ref{fig:krech_knc_Qv_Pc} and \ref{fig:krech_knv_Qc_Pc} , showing the radial profiles of the Kretschmann scalar $\mathcal{K}(r)$.

Figure {\ref{fig:fh2_ge_0_P_constant} displays the radial profiles of the metric potential $f(r)h^2(r)$ for the cases $\kappa>0$ and $\kappa<0$. As seen from figures \ref{fig:fh2_kpc_Qv_Pc} and \ref{fig:fh2_kv_Qc_Pc}, $f(r)h^2(r)\rightarrow0$  as $r\rightarrow0$, in agreement with the asymptotic analysis giving equation (\ref{eq_fh2_r0_1}) for $\kappa>0$.

Figure \ref{fig:fh2_kpc_Qv_Pc} displays the profiles of $f(r)h^2(r)$ with $\kappa=3M^2$ for different values of $Q/M$. Equation (\ref{eq_fh2_sing}) indicates that $f(r)h^2(r)$ diverges as $r\rightarrow a$ (or $x\rightarrow0$), where $a=\left(\kappa Q^{2}\right)^{1/4}$. This behaviour is evident in figure \ref{fig:fh2_kpc_Qv_Pc} for the cases $Q/M=0.5$, $0.7$ and $1.2$. Likewise, figure \ref{fig:fh2_kv_Qc_Pc} displays the profiles of  $f(r)h^2(r)$ with $Q=0.5M$ for different positive values of $\kappa$. The diverging nature of $f(r)h^2(r)$ as $r\rightarrow a$ is also evident from figure \ref{fig:fh2_kv_Qc_Pc} for the cases $\kappa/M^2=3$, $10$ and $15$.

Equation (\ref{eq_fh2_sing_kn}) implies that $f(r)h^2(r)\rightarrow-\infty$ as $r\rightarrow b$ (or $x\rightarrow0$), where $b=\left(|\kappa| Q^{2}\right)^{1/4}$. This nature of $f(r)h^2(r)$ is evident from figure \ref{fig:fh2_knc_Qv_Pc} and \ref{fig:fh2_knv_Qc_Pc} for all cases of $\kappa<0$.

%==================================================
\begin{figure}[t!]%[htbp] % Adjust placement using [h]ere, [t]op, [b]ottom, [p]age
\centering
\begin{subfigure}{0.49\textwidth} % Adjust width as needed
\includegraphics[width=\textwidth, height=6cm]{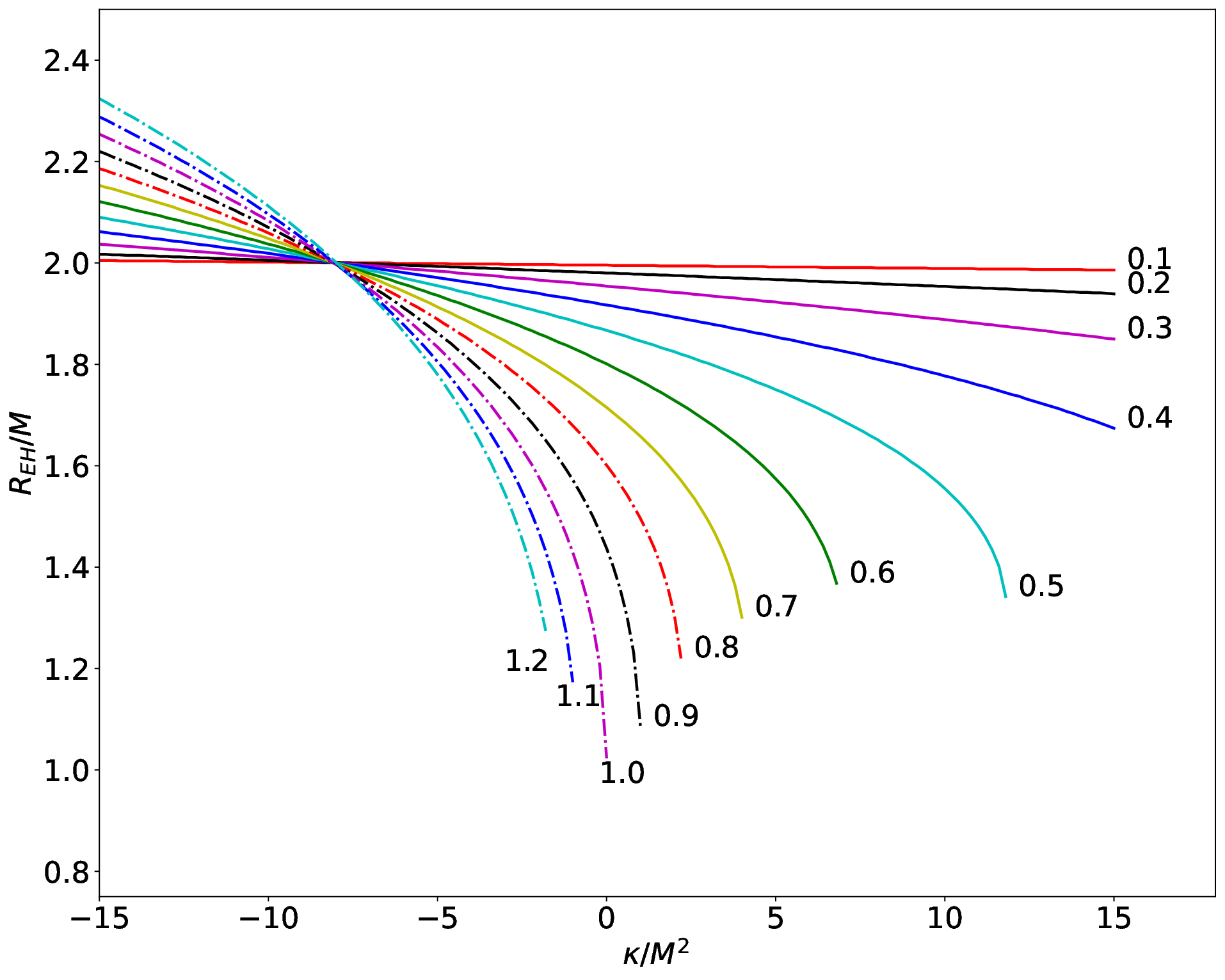} % Replace with your image path
\caption{Event horizon $R_{\rm EH}/M$ with respect to $\kappa/M^2$ for different values of the parameter $Q/M$.}
\label{fig:Reh1}
\end{subfigure}
\hfill % Space between subfigures
\begin{subfigure}{0.49\textwidth}
\includegraphics[width=\textwidth, height=6cm]{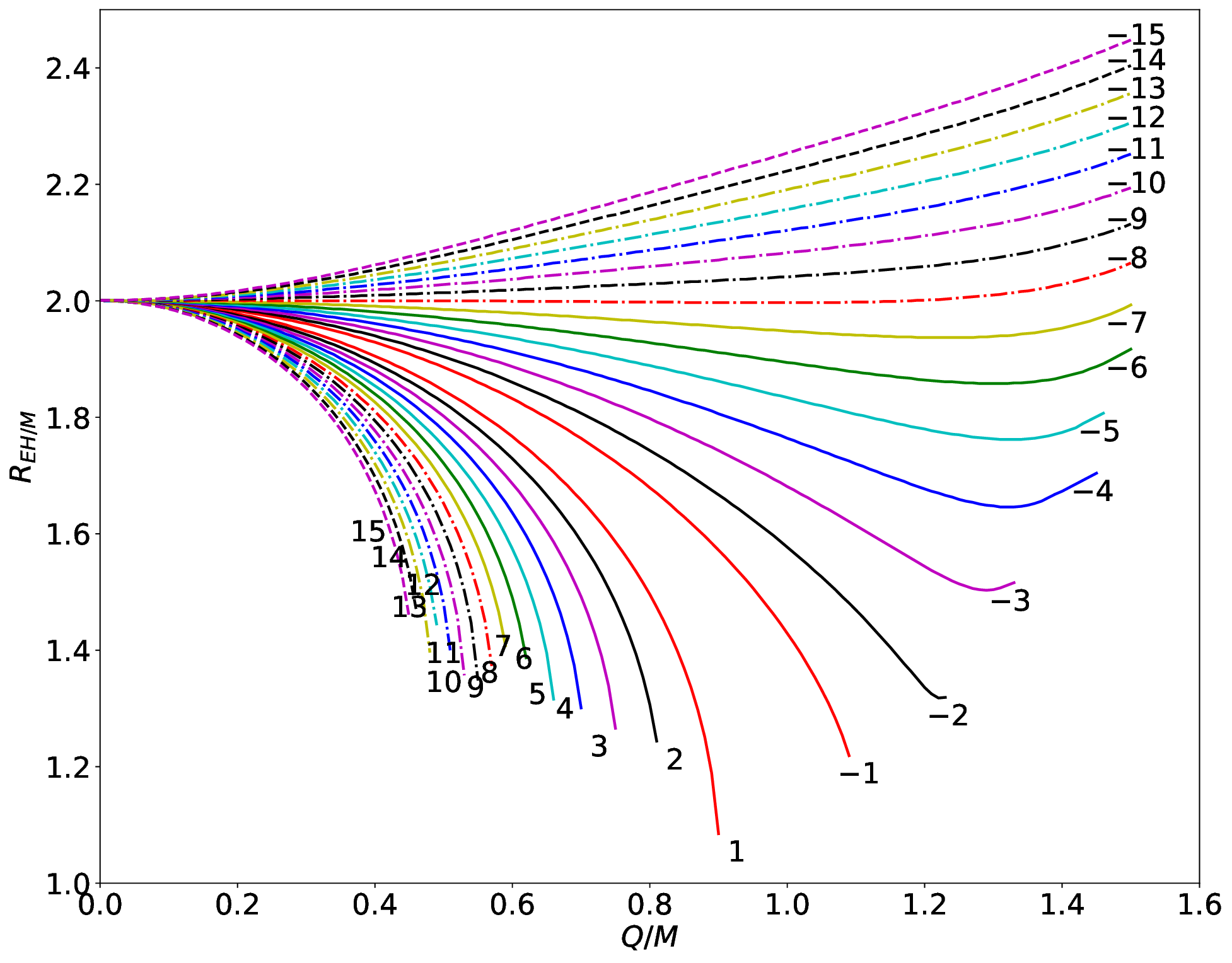}
\caption{Event horizon $R_{\rm EH}/M$ with respect to $Q/M$ for different values of the parameter $\kappa/M^2$.}
\label{fig:Reh2}
\end{subfigure}
\caption{Parametric plots for the event horizon $R_{\rm EH}/M$.}
\label{fig:EH}
\end{figure}

%==================================================

Figure \ref{fig:EH} displays the dimensionalised event horizon $R_{\rm EH}/M$ for different values of the parameters $\kappa/M^2$ and $Q/M$ for a few cases where covered black hole singularities exist.

Figure \ref{fig:Reh1} displays the parametric plot  for $R_{\rm EH}/M$ with respect to the coupling $\kappa/M^2$ for different values of the parameter $Q/M$. We see that, larger the value of the coupling $\kappa/M^2$, lesser is the amount of charge that the black hole can accommodate. When the EiBI coupling is negative, $\kappa<0$, an EiBI black hole can  accommodate charge $Q>M$, unlike the Reissner-Nordstr\"om case.

Figure \ref{fig:Reh2} displays the parametric plot  for $R_{\rm EH}/M$ with respect to $Q/M$ for different values of the parameter $\kappa/M^2$. We once again see that, when the coupling is negative, the black hole can  accommodate charge $Q>M$, unlike the Reissner-Nordstr\"om case.

\section{Discussion and conclusion}\label{sec_disc_con}

In this paper, we considered the behaviour of the spacetime of charged black holes described by Eddington-inspired Born Infeld gravity, often abbreviated as EiBI gravity. We first obtained the field equations in the Palatini formalism from the EiBI action (\ref{eq_action}) with a Maxwell field minimally coupled to gravity. We solved the field equation with a static and spherically symmetric metric leading to a differential equation for the metric coefficient $f(r)$, given by equation (\ref{eq_df}).

In order to explore the behaviour of the spacetime of the charged EiBI black hole, we carried out an in-depth analysis for asymptotic nature of the spacetime, such as its behaviour at long distances, near the center, in the intermediate region, and near-horizon behaviour, as discussed in Section \ref{sec_asymp}. In these regions of the spacetime, we analysed the behaviour of the metric coefficients $f(r)$, $h(r)$, $f(r)h^2(r)$, and the Kretschmann invariant $\mathcal{K}(r)$, that gave a thorough analytical understanding of the nature of spacetime.

We thus made, for the first time, a critical study of the behaviour of the geometry in several regions of the black hole spacetime in EiBI gravity by {\em analytical} means.

In order to aide our understanding further, we solved the differential equation (\ref{eq_df}) for the metric coefficient $f(r)$ numerically. In consequence, we obtained how the metric coefficients $f(r)$, $h(r)$, $f(r)h^2(r)$, and the Kretschmann invariant $\mathcal{K}(r)$ behave for different values of the black hole charge $Q$ and the EiBI coupling $\kappa$.

The key points of the findings from our {\em analytical} study  are the following:

\begin{itemize}
 \item The long distance behaviour of the EiBI charged black hole approaches the Reissner-Nordstr\"om spacetime with vanishing Kretschmann curvature.

 \item With a positive EiBI coupling, $\kappa>0$, the metric coefficient $f(r)$ diverges like $r^{-2}$ at the center. Although the metric coefficient $h(r)$ vanishes like $r^2$, the Kretschmann scalar diverges like $r^{-8}$ for $\kappa>0$.

 \item For a negative EiBI coupling, $\kappa<0$, the spacetime ceases to exist in the region at and around the center since the metric coefficient has imaginary parts.

 \item For the case $\kappa>0$, the metric coefficients $f(r)$ and $f(r)h^2(r)$ both diverge like $(r-a)^{-1}$, where $a=\left(\kappa Q^{2}\right)^{1/4}$, whereas the Kretschmann scalar $\mathcal{K}(r)$ diverges like $(r-a)^{-6}$ as $r\rightarrow a$.

 \item For the case $\kappa<0$, the metric coefficients $f(r)$ goes to zero like $(r-b)^{1/2}$, and $h(r)$ and $f(r)h^2(r)$ both diverge like $(r-b)^{-1/2}$, where $b=\left(|\kappa| Q^{2}\right)^{1/4}$. The Kretschmann scalar $\mathcal{K}(r)$ diverges like $(r-b)^{-6}$ as $r\rightarrow b$ in this case as well.

 \item Near the horizon, the metric coefficients and the Kretschmann scalar are finite.

\end{itemize}

We found close agreement between the above analytical behaviours and those obtained from numerical integration of the differential equation for $f(r)$ given by \ref{eq_df}. For illustration, the numerical data are displayed by several plots in Section \ref{sec_num}.

We may therefore conclude that a charged black hole in EiBI gravity, although having a intricate structure of the spacetime geometry, can lead to an in-depth understanding of the spacetime structure by means of probing the asymptotic behaviour in different regions of the spacetime by analytical means.

%\pagebreak

\section*{Acknowledgments}
Muhammed Shafeeque is supported by the Ministry of Education, Government of India, through a Research Fellowship. The Authors would like to thank the Indian Institute of Technology Guwahati for providing access to computing and supercomputing facilities.

\providecommand{\href}[2]{#2}\begingroup\raggedright\endgroup

\end{document}